\documentclass[aps, pre,twocolumn,floatfix, superscriptaddress,showpacs,longbibliography]{revtex4-2}
\pdfoutput=1 
\usepackage{amsmath,amssymb,bm}
\usepackage{graphics}
\usepackage{epsfig}
\usepackage{graphicx}
\usepackage{verbatim}
\usepackage{xcolor}
\newcommand{\be}{\begin{equation}}
\newcommand{\ee}{\end{equation}}
\newcommand{\bea}{\begin{eqnarray}}
\newcommand{\eea}{\end{eqnarray}}
\newcommand{\avg}[1]{\langle{#1}\rangle}
\newcommand{\Avg}[1]{\left\langle{#1}\right\rangle}

\newcommand{\eq}[1]{(\ref{#1})}

\graphicspath{{./_Figures/}}

\begin{document}

\title{The nature of hypergraph $k$-core percolation problems}
\author{Ginestra Bianconi}

\affiliation{
School of Mathematical Sciences, Queen Mary University of London, London, E1 4NS, United Kingdom}
\affiliation{The Alan Turing Institute,  96  Euston  Road,  London,  NW1  2DB,  United  Kingdom}

\author{Sergey N. Dorogovtsev}
\affiliation{Departamento de F\'{\i}sica da Universidade de Aveiro $\&$ I3N, 3810-193, Aveiro, Portugal} 
\affiliation{Ioffe Physico-Technical Institute, 194021 St. Petersburg, Russia}

%%\date{\today}

\begin{abstract} 
 {Hypergraphs are higher-order networks that capture the interactions between two or more nodes. Hypergraphs can always be represented by factor graphs, i.e. bipartite networks between nodes and factor nodes (representing groups of nodes). Despite this universal representation,  here we reveal that $k$-core percolation on hypergraphs can be significantly distinct from $k$-core percolation on factor graphs.  We formulate the theory of  hypergraph $k$-core percolation based on the assumption that a hyperedge can only be intact if  all its nodes are intact.} This scenario applies for instance to supply chains where  the production of a product requires all raw materials and all processing steps; in biology it applies to  protein-interaction networks where protein complexes can only function if all the proteins are present, and it applies as well to chemical reaction networks where a chemical reaction can take place only when all the reactants are present. Formulating a message-passing theory for hypergraph $k$-core percolation, and combining it with the theory of critical phenomena on networks we demonstrate sharp differences  with previously studied factor graph $k$-core percolation processes where it is allowed for hyperedges to have one or more damaged nodes and still be intact.
 {To solve the dichotomy between $k$-core percolation on hypegraphs and  on factor graphs,  we define a set of pruning processes that act either exclusively on nodes or exclusively on hyperedges and depend of their second-neighborhood connectivity.} We  show that the resulting second-neighbor $k$-core percolation problems are significantly distinct from each other. Moreover  we reveal that although these processes remain distinct from factor graphs $k$-core processes,  when the pruning process acts exclusively on hyperedges the phase diagram is reduced to the one of factor graph $k$-cores.
\end{abstract}

%\pacs{}
%\keywords{}


\maketitle

%%%%%%%%%%%%%%%%%%%%%%%%
%%%%%%%%%%%%%%%%%%%%%%%%
%%%%%%%%%%%%%%%%%%%%%%%%

%%%%%%%%%%%%%%%%%%%%%%%%%%%%%%%%%%%%%%%%
%%%%%%%%%%%%%%%%%%%%%%%%%%%%%%%%%%%%%%%%

%%%%%%%%%%%%%%%%%%%%%%%%%%%%%%%%%%%%%%%%
%%%%%%%%%%%%%%%%%%%%%%%%%%%%%%%%%%%%%%%%

%%%%%%%%%%%%%%%%%%%%%%%%
%%%%%%%%%%%%%%%%%%%%%%%%
%%%%%%%%%%%%%%%%%%%%%%%%

\section{Introduction}

%%1 \cite{bianconi2021higher} 
%%6 \cite{bianconi2018multilayer} 
%%\cite{lee2023k} 
%%\cite{coutinho2020covering} 
%%\cite{bianconi2018topological} 
%%3 \cite{benson2016higher} 
%%\cite{cantwell2023heterogeneous} 
%%\cite{weigt2006message} 
%%\cite{newman2009random}  
%%\cite{bianconi2022statistical} 
%%31 \cite{sun2021higher} 
%%\cite{yoon2011belief} 
%%2 \cite{battiston2021physics} 
%%4 \cite{battiston2020networks} 
%%5 \cite{majhi2022dynamics} 
%%\cite{newman2023message} 
%% \cite{barthelemy2022class} 
%% \cite{boccaletti2023structure} 
%%\cite{lambiotte2018simplicial} 
%%\cite{ferraz2021phase} 
%%\cite{bianconi2018topological} 
%%32 \cite{bianconi2023theory} 
%%\cite{dorogovtsev2022nature} 
%%\cite{dorogovtsev2006k}
%%\cite{goltsev2006k} 
%%\cite{azimi2014k} 
%%\cite{serafino2020superspreading} 
%%\cite{pastor2016distinct} 

Higher-order networks---hypergraphs and simplicial complexes, representing multi-node interactions, have recently gained significant attention \cite{bianconi2021higher,battiston2021physics,benson2016higher,bianconi2018topological,battiston2020networks,majhi2022dynamics,boccaletti2023structure,ferraz2021phase,barthelemy2022class,lambiotte2018simplicial,coutinho2020covering,bianconi2022statistical}. 
Research interest is growing in both modeling higher-order network structures and investigating dynamical processes and cooperative systems on such networks \cite{sun2023dynamic,de2020social,ferraz2021phase,iacopini2019simplicial,lee2023k,mancastroppa2023hyper,millan2020explosive,st2021universal,peng2023message,bianconi2018topological,lopez2013weighted}. 
Importantly, the features and characteristics of processes and cooperative phenomena on higher-order networks differ significantly from those on ordinary networks.
In this work we focus on specific highly connected substructures in hypergraphs, namely their $k$-cores. 

Each hypergraph can be represented by an equivalent bipartite graph between nodes and hyperedges---factor nodes, which is called the factor graph of a hypergraph. 
%%, and most of theories deal with this representation.  
Considering multi-agent interactions, one can recognize two markedly distinct classes. In the first class \cite{sun2021higher,lee2023k,mancastroppa2023hyper}, the removal (damaging) of one of the interacting agents (let the number of these agents exceed $2$) doesn't break interaction between the remaining agents. Typically this happens in networks of social interactions. For instance a online social network group might still be working if one or more of its members are  not participating on it actively. Such systems of multi-node interactions are naturally described by factor graphs, i.e. bipartite networks. 
In the second class of multi-agent interactions, the removal (damaging) of one of the interacting agents breaks interaction between the remaining agents. Supply chains and catalytic networks \cite{thurner2010schumpeterian,hanel2005phase}, protein-interaction networks \cite{klimm2021hypergraphs} and networks of chemical reactions \cite{jost2019hypergraph}, provide an example for higher-order interactions of this kind. Such systems of multi-node interactions are described by hypergraphs, in which the removal of a node results in the disappearance of all the adjacent hyperedges. For instance the removal of a raw material will impede the production of a product, the absence of a protein will impede the formation of a protein complex and the absence of a reactant will impede a chemical reaction to occur. Node percolation problems for these two classes of systems qualitatively differ from each other, compare Refs.~\cite{sun2021higher} and \cite{bianconi2023theory}, although edge percolation on hypergraphs coincides with factor node percolation on corresponding factor graphs.    

The issue of $k$-cores in ordinary networks has been extensively explored \cite{seidman1983network,bollobas1984evolution,chalupa1979bootstrap,dorogovtsev2006k,goltsev2006k,dorogovtsev2022nature}. 
For a graph $G$, a $k$-core is the maximal subgraph $G_k$, in which each node has degree at least $k$, where $k$ is a given threshold value. One can decompose a graph into the set of $k$-shells $G_k\backslash G_{k+1}$ and classify the nodes according to shells to which they belong \cite{alvarez2006large,carmi2007model}. The higher $k$-cores play a particular role in a graph in respect of rapid spreading phenomena, including fast disease spreading \cite{kitsak2010identification,serafino2020superspreading}. The highest $k$-core was observed to be the center localization in a number of network architectures \cite{pastor2016distinct}. 
The pruning process resulting in a $k$-core is quite efficient algorithmically: one must progressively remove each node with degree smaller than $k$ until no such nodes remain. 
The result of this process in an infinite graph is a subgraph which can contain a single giant and many finite components. 
Sometimes it is this giant component that is referred to as the $k$-core. We shall focus on this component. 
In the infinite tree-like networks, finite components in the $k$-cores are vanishingly rare if $k$ exceeds $1$.   

%%Hypergraphs and simplicial complexes form the important class of networksÑso-called higher-order networks
%%[1Ð5]Ñrepresenting the systems of multi-node interactions. Growing research interest is addressed to both
%%modelling [3, 6Ð8] higher-order network structure and investigating dynamical processes on top of them. Many
%%processes and cooperative models on the higher-order
%%networks significantly differ from those on ordinary networks in which each edge interconnects a pair of nodes
%%[3, 9, 10]. These include opinion dynamics, game theory,
%%synchronization etc. 

The $k$-core problems for hypergraphs have  received little attention thus far. The authors of Ref.~\cite{lee2023k,mancastroppa2023hyper} introduced the $(k,n)$-core in a factor graphs as the maximal subgraph of the factor graph with nodes of degree equal at least $k$ and factor nodes of degree equal at least $n$. 
This subgraph is the result of the progressive pruning of nodes with degrees smaller than $k$ and factor nodes with degrees smaller than $n$. This definition and the pruning algorithm applied to a factor graph are relevant for the systems that are described by bipartite networks, like social interactions mentioned earlier. If we inspect, however, hypergraphs represented by the factor graphs emerging during the execution of this algorithm, we observe that during this pruning  the cardinalities of hyperedges corresponding to factor nodes can decrease. 
Consequently, the $(k,n)$-cores introduced in this way are not subhypergraphs of the hypergraph in contrast to the $k$-cores of an ordinary graph.

 {In this work we adopt definitions of $k$-core hypergraph percolation} that pertain to specific multi-node interactions in which the removal or damage of one interacting agent disrupts the interaction among the remaining agents (e.g., supply chain, protein interaction networks, networks of chemical reactions). 
In the present work 
%%we describe a set of $k$-core problems for hypergraphs introduced in the traditional way, similar to .  
we describe a set of $k$-core problems specific for hypergraphs and the corresponding pruning algorithms in which nodes and hyperedges are progressively removed (damaged) if their degrees and cardinalities, respectively, are smaller than given threshold values, $k$ and $n$, and hence each step of these algorithms provides a subhypergraph of an initial hypergraph. 
The $(k,n)$-cores produced by these pruning algorithms are the maximal subhypergraphs whose vertices and hyperedges have degrees and cardinalities equal at least $k$ and $n$. 
These definitions specifically address multi-node interactions that exhibit the following characteristic: if one of the interacting agents is removed, it disrupts the interaction among the remaining agents (as in chemical reactions).

For such $(k,n)$-cores in random hypergraphs, we explore phase transitions and obtain phase diagrams. 
These phase diagrams are significantly more rich than for the $k$-cores in ordinary graphs,  {and in factor graphs  \cite{lee2023k,mancastroppa2023hyper}} where the phase transition for the $2$-core is continuous, while the phase transitions for $(k\geq3)$-cores are hybrid.
% i.e. discontinuous transitions displaying a singularity above the transition  {(see for definition of hybrid transitions and background information~\cite{dorogovtsev2008critical,dorogovtsev2006k,bianconi2018multilayer}).}
 {In particular, we observe a significant difference between the  critical properties of the  $(2,2)$-core on factor graphs and on hypergraphs.}  {While in factor graphs the  $(2,2)$-core percolation is always continuous \cite{lee2023k,mancastroppa2023hyper}, on hypergraphs} we  observe two transition lines on the phase diagram---the continuous transition line and the hybrid transition one. These lines converge at the tricritical point.  

 {In order solve the dichotomy between $(k,n)$-core percolation on factor graphs \cite{lee2023k,mancastroppa2023hyper} and the $k$-core  percolation on hypergraphs investigated here } we introduce a novel class of  $k$-core problems, in which the pruning process accounts not only for the closest neighborhood of a node (e.g., it accounts not only for hyperedges adjacent to a node but also for all their nodes). 
In this class of models, 
%%, in contrast to ordinary networks, 
the pruning process can involve either exclusively nodes or exclusively hyperedges and the pruning algorithm might depend on the nodes/hyperedges which are second neighbors within the factor graph.  

We show that the second-neighbor $k$-core percolation involving exclusively pruning of the nodes or involving exclusively pruning of the hyperedges are distinct, and we relate these models to both hypergraph $(k,n)$-cores depending on the closest neighborood and to factor graphs $(k,n)$-core.  In particular, we show that the percolation threshold for the second-neighbor $k$-core problems with pruning of the nodes coincides with the percolation threshold for the first-neighbor hypergraph $k$-core problems, however in the limiting case in which only hyperedges are initially damaged, the second-neighbor $k$-core problems with pruning of the hyperedges coincides with the percolation threshold of the factor graph $k$-cores  {\cite{lee2023k,mancastroppa2023hyper}}.

These results are obtained within a message-passing theory \cite{newman2023message,weigt2006message,bianconi2018multilayer,cantwell2023heterogeneous} exact for locally tree-like hypergraphs and within the  {generating function } theory of critical phenomena on networks and it is here supported by simulations. 
 {The message-passing equations for the $k$-core percolation problems presented here are derived  from their definition of the $k$-cores problems using as starting point the message-passing for percolation in hypergraphs  \cite{bianconi2023theory}.} 

Our approach is 
%%very 
general, and the message-passing equations can be applied to arbitrary locally tree-like hypergraphs. 
In particular, we apply these equations 
%%Here we focus, in particular, to the application 
to random hypergraphs \cite{bianconi2022statistical,newman2009random,yoon2011belief} with given cardinality and degree distributions.
%%These results are obtained within a message-passing theory \cite{newman2023message,weigt2006message,bianconi2018multilayer} exact for locally tree-like hypergraphs and within the theory of critical phenomena on networks supported by simulations. 
Possibly the proposed approach could be extended in the future in order to go beyond the locally tree-like approximation thanks to recent advances on message passing on networks with loops \cite{kirkley2021belief,cantwell2023heterogeneous}.

The paper is structured as follows: 
In Sec.~\ref{s2} we overview the $k$-core problem for ordinary graphs and develop the message-massing theory for it.  
Section~\ref{s7} introduces the basic definitions and notations for hypergraphs, in particular, the definition of the  subhypergraph of a hypergraph. 
In Sec.~\ref{s4p}  {we derive the message-passing and the generating functions equations} for the $(k,n)$-core problem on hypergraphs (the first-neighbor version in the sense of nodes and factor nodes in factor graphs). 
In Sec.~\ref{second_neighbor}  {we derive the message-passing and the generating functions equations} for  the second-neighbor version of the $(k,n)$-core problem on hypergraphs.  
In Sec.~\ref{s-c} we provide the concluding remarks.

%%%%%%%%%%%%%%%%%%%%%%%%%%%%%%%%%
%%%%%%%%%%%%%%%%%%%%%%%%%%%%%%%%%
%%%%%%%%%%%%%%%%%%%%%%%%%%%%%%%%%

\section{Overview of $k$-core percolation on simple networks}
\label{s2}

\subsection{$k$-core and pruning algorithm}

We consider a graph $G=(V,E)$. 
%%Let us consider an initial random damage of the nodes. 
The $k$-core is the largest subgraph where intact vertices have at least $k$ interconnections. We start from a configuration in which nodes are initially damaged  with probability $1-p$.
The $k$-core is obtained by the following algorithm: 
\begin{itemize}

\item[(1)] Damage iteratively all the nodes with less than $k$ undamaged neighbors.

\item[(2)] The $k$-core is the giant component of the network induced by the undamaged nodes. This $k$-core is the giant subgraph induced by  nodes left undamaged by the pruning process.

\end{itemize}
Note that in infinite 
%%random 
locally tree-like 
%%random 
graphs finite $k$-core components  
%%formed by undamaged nodes  
are negligible.

\subsection{Derivation of the message-passing algorithm for $k$-cores}

%%The equations for $k$-core on a random graphs  are typically derived by observing that the $k$-core coincides with the infinite $(k{-}1)$-ary subtree.
Here we aim to derive the message-passing equations for the $k$-core starting directly from the pruning algorithm under the hypothesis that the network is locally tree-like. 
To this end, let us assume that the initial damage of the nodes is exactly known and encoded in the indicator function $x_i\in \{0,1\}$ specifying whether a node is initially damaged $x_i=0$ or not $x_i=1$.

At step (2) we assume to know whether each node $i$ has been pruned/damaged ($s_i=0$) or not ($s_i=1$).  The message-passing equations determining the giant component of the network formed by undamaged nodes are:
\bea
\hat\sigma_{i\to j}=s_i\left[1-\prod_{r\in N(i)\setminus j}(1-\hat\sigma_{r\to i})\right],
\label{p_m0}
\eea
 {where $N(i)$ denotes the set of neighbors of node $i$.}
Note that the message $\hat\sigma_{i\to j}$ indicates whether node $j$ connects node $i$ to the giant component ($\hat\sigma_{i\to j}=1$) or not ($\hat\sigma_{i\to j}=0$) and it is defined assuming that node $j$ is already in the giant component (see for instance discussion of the message-passing algorithm in multilayer networks with link overlap \cite{cellai2016message,radicchi2017redundant,bianconi2018multilayer}).
The indicator function determining whether node $i$ belongs to the giant component or not is instead given by 
\bea
\hat\sigma_i=s_i\left[1-\prod_{r\in N(i)}(1-\hat\sigma_{r\to i})\right].
\label{per_mess}
\eea
 
Now at step (2) we have that a node  that is undamaged and belongs to the $k$-core must receive  at least $k$ positive messages (i.e. it should be connected to at least $k$ nodes in the giant component), i.e.
\bea
s_i=x_i
 \sum_{\substack{\Theta \subseteq N(i) \\ |\Theta| \geq k}}\ \ \prod_{r \in \Theta} \hat\sigma_{r \to i}\prod_{r \in N(i)\setminus \Theta} (1 - \hat\sigma_{r \to i}),
\label{0100}
\eea
where $\Theta$ is a subset of $N(i)$ including at least $k$ nodes.
Under the assumption that $j$ is connected to the giant component,i.e. $\hat{\sigma}_{j\to i}=1$, substituting the above expression for $s_i$ into Eq.~(\ref{p_m0})  for $\hat{\sigma}_{i\to j}$ we obtain then that at stationarity the messages  $\sigma_{i\to j}$ satisfy
\bea
\hspace{-17pt}\hat\sigma_{i\to j}=x_i \sum_{\substack{\Theta \subseteq N(i)\setminus j \\ |\Theta| \geq k-1}}\ \ \prod_{r \in \Theta} \hat\sigma_{r \to i}\prod_{r \in N(i)\setminus (\Theta\cup j)} (1 - \hat\sigma_{r \to i})
,  
\label{0900}
\eea
while $\hat{\sigma}_i=s_i$ is given by Eq.~(\ref{0100}).

These message-passing equations can be averaged over the distribution of $\{x_i\}$ given by 
\bea
P(\{x_i\}) = \prod_{i=1}^Np^{x_i}(1-p)^{1-x_i}
,
\eea 
and using the locally tree-like approximation we get 
\bea
\!\!\! \!\!\! \!\!\! 
\sigma_{i\to j} &=& p
 \sum_{\substack{\Theta \subseteq N(i)\setminus j \\ |\Theta| \geq k-1}}\ \ \prod_{r \in \Theta} \sigma_{r \to i}\prod_{r \in N(i)\setminus (\Theta\cup j)} (1 - \sigma_{r \to i})
 ,
\label{010a}
\\[3pt]
\!\!\! \!\!\! \!\!\! 
\sigma_i &=& p\sum_{\substack{\Theta \subseteq N(i) \\ |\Theta| \geq k}}\ \ \prod_{r \in \Theta} \sigma_{r \to i}\prod_{r \in N(i)\setminus \Theta} (1 - \sigma_{r \to i})
.
\label{010b}
\eea

Since these message-passing equations are more cumbersome to implement than the original pruning process, typically the message-passing algorithms are not applied to the $k$-core problems. However their formulation can be used to derive 
%%the exact formula of 
the 
%%analytic 
equations 
%%capturing 
describing the pruning algorithm on a random  (locally tree-like) graph that we discuss in the next paragraphs. 
%Furthermore this approach has the advantage that can be directly extended to treat different version of hypergraph $(k,n)$-cores as we will see in the following sections.  

\subsection{ The $k$-core transition on  ordinary random networks}
\label{s8}

Having derived the message-passing algorithm describing the final outcome of the pruning process, we can now demonstrate how the known formulas for $k$-core percolation on a random network relate to the pruning algorithm. 
%%As we will see in the following paragraphs of this paper this 
This exercise will help us clarify the correct equations determining $(k,n)$-core percolation on hypergraphs. 

Let us 
%%then 
consider the $k$-cores of networks 
%%generated using  
provided by the configuration model with a given degree distribution $P(q)$.
In the configuration model, to each network $G=(V,E)$ with $N=|V|$ nodes and the adjacency matrix ${\bf a}$, the following probability is assigned 
\bea
P(G)=\prod_{i<j}p_{ij}^{a_{ij}}(1-p_{ij})^{1-a_{ij}}
\eea
with 
\bea
p_{ij}=\frac{q_iq_j}{\avg{q}N}
, 
\eea
where $q_i$ is the degree of node $i$.

By averaging the message-passing equations over a network generated by the configuration model we get  that $Z$ indicating the average message $\sigma_{i\to j}$ in the configuration network ensemble  is given by 
\be
Z = \sum_{q=k}^\infty \frac{q P(q)}{\avg{q}} \sum_{s=k-1}^{q-1} {q-1 \choose s} Z^s (1 - Z)^{q-1-s}
,
\label{300}
\ee
and that $S_k = \Avg{\sigma_i}$ indicating the fraction of nodes in the $k$-core, is given by 
\bea
S_k = \sum_{q=k}^\infty P(q) \sum_{s=k}^q {q \choose s} Z^s (1 - Z)^{q-s}.
\label{300Sk}
\eea
This latter quantity can be also written as 
\be
S_k = \sum_{s=k}^\infty S_k(s)
.
\label{310}
\ee
where $S_k(s)$ is the fraction of nodes that are in the $k$-core and have exactly degree $s\geq k$ within it.
We have 
\be
S_k(s) = \sum_{q=s}^\infty P(q) {q \choose s} Z^s (1 - Z)^{q-s}
,
\label{320}
\ee
and Eq.~(\ref{310}) follows from the observation that 
\bea
S_k &=& \sum_{q=k}^\infty P(q) \sum_{s=k}^q {q \choose s} Z^s (1 - Z)^{q-s} 
\nonumber
\\[3pt]
&=& \sum_{s=k}^\infty \sum_{q=s}^\infty P(q) {q \choose s} Z^s (1 - Z)^{q-s}
,
\label{330}
\eea
where we have used the equality: 
\be
\sum_{q=k}^\infty \sum_{s=k}^q = \sum_{s=k}^\infty \sum_{q=s}^\infty
.
\label{340}
\ee
We conclude our overview of $k$-core percolation on an ordinary network by expressing Eq.~(\ref{300}) and Eq.~(\ref{300Sk}) in terms of the generating functions of a degree distribution, $G(z) \equiv \sum_q P(q) z^q$, getting 
\be
Z = 1 - \frac{1}{\avg{q}} \sum_{s=0}^{k-2} \frac{Z^s}{s!} G^{(s+1)}(1 - Z)
%%.
\label{350}
\ee
and 
\be
S_k = 1 - \sum_{s=0}^{k-1} \frac{Z^s}{s!} G^{(s)}(1 - Z)
.
\label{360}
\ee
%%

%%%%%%%%%%%%%%%%%%%%%%%%
%%%%%%%%%%%%%%%%%%%%%%%%
%%%%%%%%%%%%%%%%%%%%%%%%
%%%%%%%%%%%%%%%%%%%%%%%%
%%%%%%%%%%%%%%%%%%%%%%%%
%%%%%%%%%%%%%%%%%%%%%%%%

\section{Subhypergraphs of hypergraphs}
\label{s7}

Denote a hypergraph by $\mathcal{H} = (V,H)$, where $V$ and $H$ are the sets of its vertices and hyperedges. We indicate with $N$ the number of nodes, i.e. $|V|=N$, and with $M$ the number of hyperedges, i.e. $|H|=M$, of the hypergraph.
We 
%%will 
indicate the nodes of the hypergraph with Latin letters $i,j,r,\ldots$ and the hyperedges of the hypergraph with Greek letters $\alpha,\beta,\gamma,\ldots$.
Each hyperdege $\alpha$ determines a set of nodes 
\bea
\alpha=[i_1,i_2,i_3,\ldots i_{m_\alpha}]
\eea
with $m_\alpha \equiv |\alpha|$ indicating the cardinality (number of nodes) of the hyperdege $\alpha$.
Likewise, each node $i$ has degree $q_i$ indicating the number of hyperedges it belongs to.

A subhypergraph $\mathcal{S}$ of the hypergraph $\mathcal{H}$ 
%%$by $\mathcal{H}$ 
is defined as $\mathcal{S} = (V_S,H_S)$, where $V_S \subset V$ and $H_S \subset H_\text{max}$, where $H_\text{max}$ is the full set of those hyperedges of $H$ that have each of their end vertices belonging to $V_S$. 
In particular, $\mathcal{S}_\text{induced} = (V_S,H_\text{max})$ is the vertex induced subhypergraph of the hypergraph $\mathcal{H}$, induced by the set of vertices $V_S$. 
Importantly, any subhypergraph $S$ of the hypergraph $\mathcal{H}$ cannot have hyperedges not belonging to $\mathcal{H}$, for example, hyperedges of smaller cardinalities.

%%%%%%%%%%%%%%%%%%%%%%%%%%%%%%%%
%%%%%%%%%%%%%%%%%%%%%%%%%%%%%%%%
%%%%%%%%%%%%%%%%%%%%%%%%%%%%%%%%

\section{Hypergraphs $(k,n)$-core problems  and their (first-neighbor) pruning algorithm}
\label{s4p}
\begin{figure*}[tbh!]
%%[htbp]
\begin{center}
\includegraphics[width=1.8\columnwidth]{ 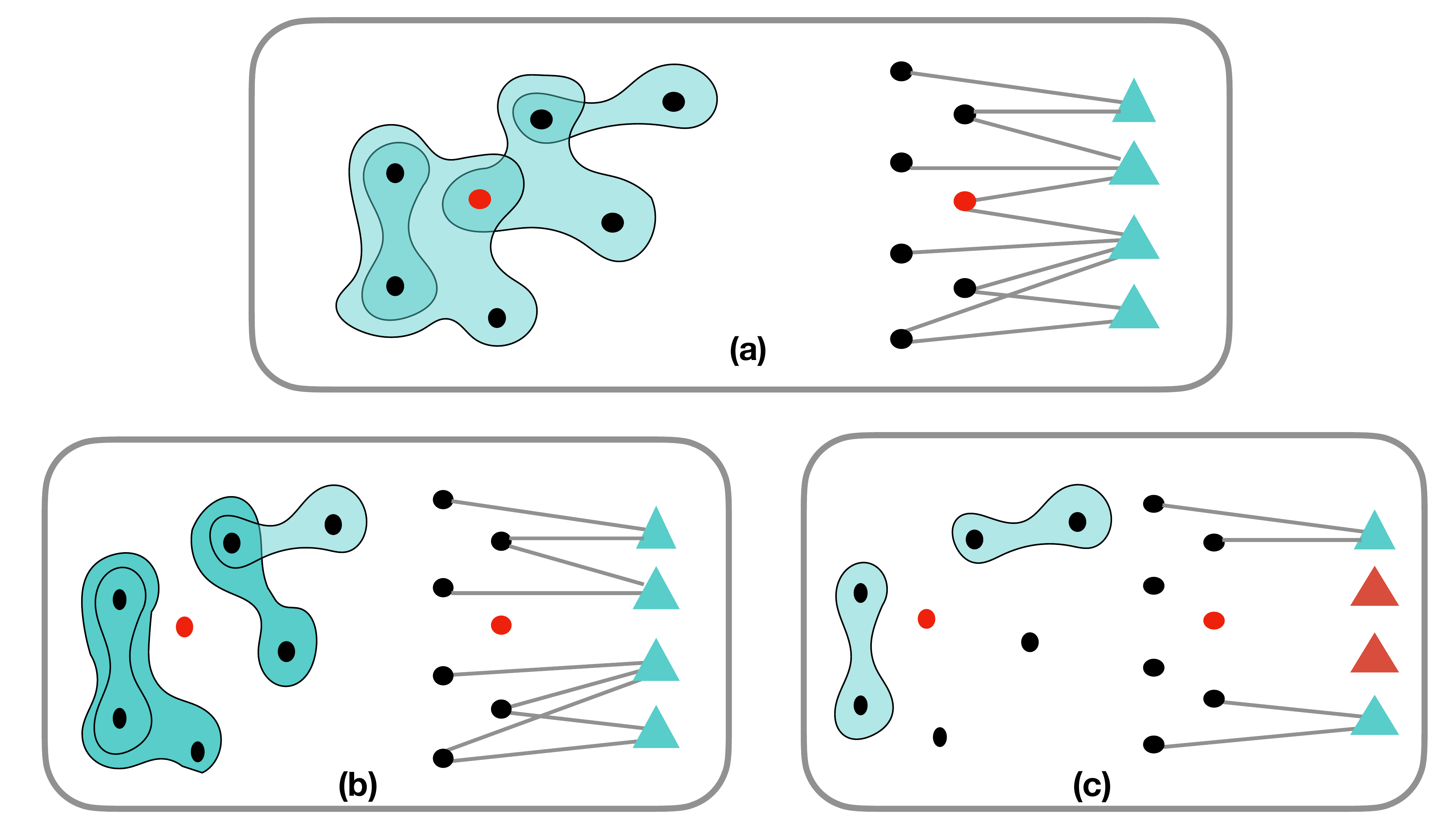}
\end{center}
\caption{Schematic representation of the difference between the factor-graph \cite{lee2023k,mancastroppa2023hyper} and hypergraph pruning algorithms. We consider the hypergraph in panel (a) and its factor graph representation. In the factor-graph pruning algorithm (panel (b)), when a node is damaged, all the hyperedges including this node are reduced in size by one. In hypergraph percolation (panel (c)) all hyperedges containing the damaged node are damaged. 
%%From the figure it is immediate to appreciate 
The figure demonstrates that only the hypergraph pruning algorithm generates a subhypergraph of the original hypergraph.}
\label{f1}
\end{figure*}

\subsection{The first-neighbor pruning algorithm}
\label{first_neighbor}

We consider a hypergraph $\mathcal{H} = (V,H)$. In this hypergraph we 
%%will 
assume that a hyperedge is  intact if it is not damaged and if none of its nodes are damaged.
The hypergraph can be always represented as a factor graph which is a bipartite networks having two types of nodes, namely, the nodes corresponding to the nodes of the hypergraphs and the factor nodes corresponding to the hyperedges of the hypergraph.
Accordingly, we can choose to iteratively prune only nodes, only  hyperedges, or both nodes and hyperedges. 
%%are iteratively pruned.
If the algorithm for the pruning only depends on the first neighbors of a node in the factor graph, we can treat all these variants simultaneously. 
On the other hand, when the pruning algorithm depends on the first and second neighbors of a node in the factor graph,  we need to treat independently pruning of nodes and pruning of hyperedges as we will show in Sec.~\ref{second_neighbor}. 
Moreover, for each of the pruning algorithms, the initial damage can target either nodes or hyperedges.

If we consider the pruning on the nodes and hyperedges depending only of the state of their neighbors in the factor graph we can define the hypergraph's $(k,n)$-core.
 The $(k,n)$-core is the maximal subhypergraph with the vertices whose (internal) degrees are at least $k$ and the hyperedges have cardinalities equal or exceeding $n$. 
We start  from a configuration in which nodes are initially damaged with probability $1-p_N$ and/or hyperedges are initially damaged with probability $1-p_H$.
The $(k,n)$-core can be  obtained  using  the following  pruning algorithm: 
\begin{itemize} 

\item[(1)] Damage iteratively all the hyperedges  having less than $n$ (undamaged) nodes and all the nodes with less than $k$ undamaged hyperedges. 

\item[(2)] The $(k,n)$-core is the giant subhypergraph  induced by undamaged nodes and undamaged hyperedges. 

\end{itemize}
Equivalently, this $(k,n)$-core can be obtained by, first, removing all hyperedges with cardinalities smaller than $n$ and,  second, progressively pruning all nodes with degrees smaller than $k$. 
The $(k,2)$-core can be called the $k$-core for hypergraphs. 

 {From the above definition of hypegraph $k$-core percolation we conclude that there are two major differences between hypegraph percolation  and factor graph percolation \cite{lee2023k,mancastroppa2023hyper}. First, and most importantly, in hypergraph percolation  the damage of a node automatically disrupts all the hyperedges to which the node belongs while on factor graph percolation only reduces by one the degree of the factor nodes to which the node is connected. Secondly hypergraph $k$-cores are subhypegraphs of the original hypergraph while this property is not preserved in factor graph percolation (see schematic representation of the difference   in Fig.~\ref{f1}). }
%%

%%
%We note that node and hyperedge percolation are distinct as outlined in Ref.~\cite{bianconi2023theory} and that a hyperedge is in the giant component only if (i) it is not initially damaged; (ii) none of its nodes are initially damaged; (iii) at least one of its nodes is connected to the giant component.
%If follows that  the above algorithm is equivalent to the one in which at step (1) each hyperedge $\alpha$ is damaged if and only if its cardinality is smaller than $n$, i.e. $|\alpha|<n$ and nodes are progressively pruned if they have degree smaller than $k$. 
%This is the algorithm that we will implement with message-passing equations in the following paragraph.

\subsection{Message-passing algorithm for the hypergraph $(k,n)$-core }

 {It this paragraph we derive the message-passing equations  determining the $(k,n)$-core directly from the definition of problem given in the previous paragraph.}

 {The obtained equations are very general and apply to every hypergraph under the assumption that the factor graph encoding for the hypergraph is locally tree-like.}

 {Since the definition of the $(k,n)$-core is given in terms of the giant subhypergraph induced by intact nodes and  hyperedges, the  message-passing equations for $(k,n)$-core percolation will be derived starting from the equations valid for hypergraph percolation \cite{bianconi2023theory}, which will be used to  identify this giant subhypergraph.}

 {Here we recall that on hypergraph percolation~\cite{bianconi2023theory} node and hyperedge percolation are distinct  and that an hyperedge is in the giant subhypergraph only if (i) it is not  damaged; (ii) none of its nodes are  damaged; (iii) at least one of its nodes is connected to the giant subhypergraph.}

 If follows that  the $(k,n)$ core algorithm described in the previous paragraph is equivalent to the one in which the initial damage is modified as in the following: nodes are initially damaged with probability $1-p_N$ and hyperedges $\alpha$ are initially damaged deterministically if their cardinality is smaller than $n$, i.e. $|\alpha|<n$ while if their cardinality is larger or equal to $n$, i.e. $|\alpha|\geq n$ they are damaged with probability $1-p_H$. 
Given this initial damage, the $(k,n)$ core  defined by the pruning algorithm is equivalent to the one
  obtained  using  the following pruning algorithm: 
\begin{itemize} 
\item[(1')] Damage iteratively all the nodes with less than $k$ undamaged hyperedges. 
\item[(2')] The $(k,n)$-core is the giant subhypergraph  induced by undamaged nodes and undamaged hyperedges. 
\end{itemize}

 {In order to derive the message-passing algorithm for $(k,n)$ core percolation directly from this pruning algorithm,} 
%This derivation  will 
%%make explicit use of 
%explicitly use 
%the message-passing equations recently derived for hypergraph percolation.
let us assume that the initial damage of the nodes is exactly known and encoded in the indicator function $x_i\in \{0,1\}$ indicating whether a node is initially damaged $x_i=0$ or not $x_i=1$. Similarly, we assume that the initial damage of the hyperedges is exactly known and encoded by the product $y_{\alpha}\theta(|\alpha|-n)\in \{0,1\}$ where $y_{\alpha}\in \{0,1\}$ indicates whether the hyperedge $\alpha$ is randomly damaged while $\theta(|\alpha|-n)$ enforces the deterministic damage of hyperedges of cardinality less then $n$. [Note that here $\theta(z)$ indicates the Heaviside function $\theta(z)=1$ if $z\geq 0$ and $\theta(z)=0$ otherwise.] 

The message-passing equations for $(k,n)$ core percolation are here derived starting from the definition of the $(k,n)$ core and the message-passing equation for hypergraph percolation~\cite{bianconi2023theory}.

At step (2') of the  pruning algorithm, assuming that we know the indicator functions $s_i\in \{0,1\}$ and $s_\alpha\in \{0,1\}$ indicating whether a node $i$ or a hyperedge $\alpha$ are intact or not,  {the message-passing equations that determine the nodes in the giant subhypergraph induced by the intact nodes and hyperedges are the ones of hypergraph percolation
\cite{bianconi2023theory}:}
\bea
\hat{w}_{i\rightarrow \alpha} &=&s_i \left[ 1 - \prod_{\beta\in N(i)\setminus \alpha}(1 - \hat{v}_{\beta\rightarrow i}) \right]
,
\label{hv} 
\\
\hat{v}_{\alpha\to i} &=&s_{\alpha}\left(\prod_{j\in N(\alpha)\setminus i}\!s_j\!\right)\left[ 1 -\prod_{j\in N(\alpha)\setminus i}\!(1 - \hat{w}_{j\rightarrow \alpha}) \right]
,
\nonumber
\eea
 {where $N(i)$ denotes the set of neighbors of node $i$ and $N(\alpha)$ indicates the set of neighbors of factor node $\alpha$}. 
The indicator functions $\hat\sigma_i\in \{0,1\}$ and $\hat{r}_{\alpha}\in \{0,1\}$ indicating whether nodes and hyperedges are in the giant subhypergraph  are expressed in terms of $s_i$ and $s_{\alpha}$ and are  given respectively by the equations \cite{bianconi2023theory}
\bea
\hat{\sigma}_i&=&s_i \left[ 1 - \prod_{\beta\in N(i)}(1 - \hat{v}_{\beta\rightarrow i}) \right],
\nonumber 
\\
\hat{r}_{\alpha} &=&s_{\alpha}\left(\prod_{j\in N(\alpha)}\!s_j\!\right)\left[ 1 -\prod_{j\in N(\alpha)}\!(1 - \hat{w}_{j\rightarrow \alpha}) \right]
.
\eea 
 {Note however that these equations assume that the indicator functions $s_i$ and $s_{\alpha}$ are know while here we want to obtain message-passing equations also able to determine their value as obtained by implementing the pruning algorithm (1').}
According to the definition of the hypergraph $(k,n)$-core, the  indicator function $s_i$ obtained by the pruning algorithm  is  only non-zero if the node $i$ receives at least $k$ positive messages from its neighbors, i.e. 
\bea
s_i=x_i\sum_{\substack{\Theta \subseteq N(i) \\ |\Theta| \geq k}}\ \ \prod_{\beta \in \Theta} \hat{v}_{\beta \to i}\prod_{\gamma \in N(i)\setminus \Theta} (1 - \hat{v}_{\gamma \to i})
.
\label{010a}
\eea
The indicator function $s_{\alpha}$ obtained by the pruning algorithm  is  simply given by 
\bea
s_{\alpha}=y_{\alpha}\theta(|\alpha|-n)
.
\eea
In order to get the message-passing equations for $(k,n)$ core percolation we need to insert  these expressions for $s_i$ and $s_{\alpha}$ into  Eq.~(\ref{hv}). Considering that the messages $\hat{v}_{\alpha\to i}$ is defined under the assumption that node $i$ is in the giant subhypergraph  and $\hat{w}_{i\to \alpha}$ is defined under the assumption that hyperedge $\alpha$ is in the giant subhypergraph,
 and exploiting the fact that the messages take only the $0,1$ values, we get
\bea
\hat{w}_{i\to \alpha}&=&x_i\sum_{\substack{\Theta \subseteq N(i)\setminus\alpha \\ |\Theta| \geq k-1}}\ \ \prod_{\beta \in \Theta} \hat{v}_{\beta \to i}\prod_{\gamma \in N(i)\setminus (\Theta\cup \alpha)} (1 - \hat{v}_{\gamma \to i}),\nonumber \\
\hat{v}_{\alpha\to i} &=&y_{\alpha}\theta(|\alpha|-n)\prod_{j\in N(\alpha)\setminus i}\!\hat{w}_{j\rightarrow \alpha}. 
\label{Hhva}
\eea
 {Providing an intuitive explanations of these equations  and their derivation might be instructive.
The expression for $\hat{w}_{i\to\alpha}$ in (\ref{hv}) implies that the node sends a positive message to a neighbor if it is intact ($s_i=1$) and if it receives at least a positive message from one of its hyperedges. The expression for $s_i$ in Eq.(\ref{010a}) expresses that node $i$ is intact if it not intially damaged ($x_i=1$) and if receives at least $k$-positive messages from its neighbor hyperedges. It follows that under the assumption that $\alpha$ is in the giant subhypergraph, the message $\hat{w}_{i\to\alpha}$ is equal to one, if an only if $x_i=1$ and node $i$  receives at least $k-1$ positive messages from neighbor hyperedges different from $\alpha$ as expressed by the first equation in $(\ref{Hhva})$. Similarly the equation for $\hat{v}_{\alpha\to i}$ in (\ref{hv}) implies that one hyperedge can send a positive messages only if (i) it is not initially damaged, (ii) all its nodes are intact, and (iii) it receives at least a positive message from one of its nodes. The condition that all the nodes of the hyperedge must be intact, (i.e. must have $s_i=1$) combined with the expression of $s_i$ given by Eq.(\ref{010a}) implies that every node of the hyperedge should be connected to at least other $k-1\geq 1$ intact hyperedges. This happens if and only if each of these nodes sends a positive message to the hyperedge $\alpha$ leading to the the second equation (\ref{Hhva}).}

 {Following a similar line of thought,} and exploiting  the fact that both $s_i,s_{\alpha}$ and the messages $\hat{w}_{i\to \alpha},\hat{v}_{\alpha\to i}$ are taking values $0,1$, it is also immediate to show that 
\bea
\hat{\sigma}_i&=&s_i=x_i\sum_{\substack{\Theta \subseteq N(i) \\ |\Theta| \geq k}}\ \ \prod_{\beta \in \Theta} \hat{v}_{\beta \to i}\prod_{\gamma \in N(i)\setminus \Theta} (1 - \hat{v}_{\gamma \to i}),\nonumber \\
\hat{r}_{\alpha}&=&y_{\alpha}\theta(|\alpha|-n)\prod_{j\in N(\alpha)}\!\hat{w}_{j\rightarrow \alpha}. 
\label{m1}
\eea
 {Hence Eqs.(\ref{Hhva}) and (\ref{m1}) are the message passing equations for hypergraph $(k,n)$-core percolation  when the initial random damage of  nodes, i.e. $\{x_{\alpha}\}$, and of  hyperedges i.e.$\{y_\alpha\}$, is known. } 

Another set of message-passing equations hold when we do not have direct access to the configuration of the initial damage $\{x_i\}, \{y_{\alpha}\}$ but we only know the probability that nodes and hyperedges are initially intact, i.e. $p_N$ and $p_H$ respectively. This second set of message-passing equations can be simply obtained by averaging the messages over the  initial damage distribution 
\bea
P(\{x_i\},\{y_{\alpha}\})&=&\prod_{i=1}^Np_N^{x_i}(1-p_N)^{1-x_i}\nonumber \\&&\times\prod_{\alpha=1}^{M}p_H^{y_{\alpha}}(1-p_H)^{1-y_{\alpha}}.
\label{random_xy}
\eea
In this way we  obtain the following set of message-passing equations (note that the messages $w_{i\to\alpha}$ and $v_{\alpha\to i}$ now take real values between $0$ and $1$): 
\bea
{w}_{i\to \alpha}&=&p_N\sum_{\substack{\Theta \subseteq N(i)\setminus\alpha \\ |\Theta| \geq k-1}}\ \ \prod_{\beta \in \Theta} {v}_{\beta \to i}\prod_{\gamma \in N(i)\setminus (\Theta\cup \alpha)} (1 - {v}_{\gamma \to i}),
\nonumber 
\\
{v}_{\alpha\to i} &=&p_H\theta(|\alpha|-n)\prod_{j\in N(\alpha)\setminus i}\!{w}_{j\rightarrow \alpha}. 
\label{Hp_hva}
\eea
The probability $\sigma_i$ that the node $i$  belongs to the $(k,n)$-core  and the probability $r_{\alpha}$ that the hyperedge $\alpha$ belongs to the $(k,n)$-core are given by
\bea
{\sigma_i}&=&p_N\sum_{\substack{\Theta \subseteq N(i) \\ |\Theta| \geq k}}\ \ \prod_{\beta \in \Theta} {v}_{\beta \to i}\prod_{\gamma \in N(i)\setminus \Theta} (1 - {v}_{\gamma \to i}),\nonumber \\
{r_{\alpha}}&=&p_H\theta(|\alpha|-n)\prod_{j\in N(\alpha)}\!{w}_{j\rightarrow \alpha}.\label{m2}
\eea
 {It follows that Eqs.(\ref{Hp_hva}) and Eqs.(\ref{m2}) uniquely determine the hypergraph $(k,n)$ core when we only know the probabilities $p_N$ and $p_H$ that nodes and hyperedges are initially undamaged respectively.}
The fraction $S_{kn}$ of nodes in the $(k,n)$-core  {can be expressed in terms of $\sigma_i$ and $r_{\alpha}$ given by Eqs.(\ref{m2})} as 
\bea
S_{kn}=\frac{1}{N}\sum_{i=1}^N\sigma_i
\label{MSkn1}
\eea
and the fraction $R_{kn}$ of hyperedges in the $(k,n)$-core is given by 
\bea
R_{kn}=\frac{1}{M}\sum_{\alpha=1}^Mr_{\alpha}
.
\label{MRkn1}
\eea
%%

%%%%%%%%%%%%%%%%%%%%%%%%
%%%%%%%%%%%%%%%%%%%%%%%%
%%%%%%%%%%%%%%%%%%%%%%%%

\subsection{Hypergraph $(k,n)$-core percolation on a random hypergraph }
\label{s11}

In many occasions 
%%also 
the exact structure of the hypergraph might be unknown and 
%%then 
so we need to rely on predictions based on the hypergraph ensembles from which the hypergraph is drawn.
Here we consider the ensembles of random hypergraphs $\mathcal{H}=(V,H)$ of $N=|V|$ nodes and $M=|H|$ hyperedges whose node degree distribution  is $P(q)$ and whose distribution of hyperedge cardinalities is $Q(m)$. In this ensemble the probability of a hypergraph $\mathcal{H}$ of incidence matrix ${\bf b}$ is given by 
\bea
P(\mathcal{H})=\prod_{i=1}^N\prod_{\alpha=1}^Mp_{i\alpha}^{b_{i\alpha}}(1-p_{i\alpha})^{1-b_{i\alpha}},
\eea
with 
\bea
p_{i\alpha}=\frac{q_im_{\alpha}}{\avg{q}N}
,
\eea
where $q_i$ is the degree of node $i$ and $m_{\alpha}$ is the cardinality of hyperedge $\alpha$.

We can obtain the analytical equations determining the $(k,n)$-core percolation problem in this ensemble by averaging the message-passing equation over the probability $P(\mathcal{H})$.

Let us indicate with  $V$ and $W$ respectively  the averages of the messages $v_{i\to \alpha}$ and $w_{\alpha\to i}$ over the distribution $P(\mathcal{H})$. 
We obtain then the equations of $V$ and $W$ given by 
\bea
&&
\!\!\!\!\!\!\!\! 
V = p_H\sum_{m \geq n} \frac{mQ(m)}{\avg{m}} W^{m-1} 
,
\nonumber
\\[3pt]
&&
\!\!\!\!\!\!\!\!
W = p_N \sum_{q=k}^\infty \frac{q P(q)}{\avg{q}} \sum_{s=k-1}^{q-1} {q-1 \choose s} V^s (1 - V)^{q-1-s}
.
\label{450}
\eea
Similarly  the fractions of vertices, $S_{kn}$, and hyperedges, $R_{kn}$, belonging to the $(k,n)$-core, can be obtained by Eq.~(\ref{MSkn1}) and Eq.~(\ref{MRkn1}) by averaging over $P(\mathcal{H})$ giving 
\bea
&&
R_{kn} = p_H\sum_{m \geq n} Q(m) W^m  
,
\label{455}
\\[3pt]
&&
S_{kn} =  p_N \sum_{q=k}^\infty P(q) \sum_{s=k}^q {q \choose s} V^s (1 - V)^{q-s}
.
\label{460}
\eea
In particular, setting $n=2$, we obtain the formulas for the $k$-cores in this problem. 

Using the generating functions, we get
\bea
&&
V = p_H\left[G_{Q1}(W) - \sum_{m < n} \frac{mQ(m)}{\avg{m}} W^{m-1}\right] 
,
\label{465}
\\[3pt]
&&
W = p_N - \frac{p_N}{\avg{q}} \sum_{s=0}^{k-2} \frac{V^s}{s!} G^{(s+1)}_P(1 - V)
, 
\label{470}
\eea
where $G_{Q1}(x) \equiv G'_Q(x)/G'_Q(1) = G'_Q(x)/\avg{m}$, and 
\bea
&&
R_{kn} =  p_H\left[G_Q(W) - \sum_{m<n} Q(m) W^m  \right] 
,
\label{475}
\\[3pt]
&&
S_{kn} = p_N - p_N\sum_{s=0}^{k-1} \frac{V^s}{s!} G^{(s)}_P(1 - V)
.
\label{480}
\eea
%%

%%%%%%%%%%%%%%%%%%%%%%%%
%%%%%%%%%%%%%%%%%%%%%%%%
%%%%%%%%%%%%%%%%%%%%%%%%

\subsection{The critical behavior of  $(k,n)$-core percolation on random hypergraphs}
\label{s10}

The hypergraph $(k,n)$-core percolation process has a critical behavior that differ significantly from the $k$-core percolation on simple networks and the $(k,n)$-core percolation on factor graphs.
One of the most striking properties of $(k,n)$-core percolation is the presence of discontinuous phase transitions also for $k=2$, while the $k$-core percolation on simple networks and the $(k,n)$-core percolation on factor graph are discontinuous only for $k\geq 3$.
Here we will emphasize this significant difference showing that a continuous transition in hypergraph $(k,n)$-core percolation 
%%process 
is only possible for $(k,n)=(2,2)$-core percolation 
%%processes that display tricritical points identifying the parameters 
also displaying the tricritical point at which the $(2,2)$-core percolation transition changes from continuous   {to  hybrid transitions  i.e. discontinuous transitions displaying a singularity above the transition (see for definition of hybrid transitions and background information~\cite{dorogovtsev2008critical,dorogovtsev2006k,bianconi2018multilayer}).}

We consider the $(k,n)$-core percolation transitions on random hypergraphs captured by Eq.~(\ref{450}).
By defining the functions 
\bea
f_W(W)&=&p_H\sum_{m\geq n}\frac{mQ(m)}{\avg{m}}W^{m-1}
,
\nonumber 
\\
f_V(V)&=&p_N\sum_{q\geq k}\frac{qP(q)}{\avg{q}}\!\sum_{s\geq k-1}^{q-1}\!\!\left(\begin{array}{c}q-1\\s \end{array}\right)\! V^s(1-V)^{q-1-s}
,
\nonumber 
\\ 
&&
\label{450b}
\eea
we write Eq.~(\ref{450}) as 
\bea
V=f_W(W),\quad W=f_V(V)
.
\eea
These equations can be written as 
\bea
h(V)=V-f_W(f_V(V))=0
.
\eea
According to the theory of critical phenomena \cite{bianconi2018multilayer}, we see that the lines of continuous (second order) phase transitions are determined by the conditions 
\bea
h(0)=0,\quad h^{\prime}(0)=0
;
\eea
the tricritical point is 
%%instead 
determined by 
\bea
h(0)=0,\quad h^{\prime}(0)=0,\quad h^{\prime\prime}(0)=0
;
\eea
while the lines of discontinuous (hybrid) phase transitions are determined by the equations 
\bea
h(V^{\star})=0,\quad h^{\prime}(V^{\star})=0
%%,
\label{440000}
\eea
with $V^{\star}>0$.
By direct inspection of these equations, it emerges immediately that the continuous (second order) transitions lines and the tricritical point can only occur for $(k,n)=(2,2)$. 
In particular, a second order takes place for 
%%critical points occur for 
\bea
1=p_Hp_N\frac{2Q(2)}{\avg{m}}\frac{\Avg{q(q-1)}}{\Avg{q}},
\label{second_critical}
\eea
while the tricritical point occur when Eq.~(\ref{second_critical}) is satisfied together with the following equation
\bea
\hspace{-5mm}\frac{\Avg{q(q-1)(q-2)}}{\Avg{q(q-1)}}=p_Hp_N^2\frac{6Q(3)}{\avg{m}}\left(\frac{\Avg{q(q-1)}}{\avg{q}}\right)^2.
\eea

Let us apply these equations to the uncorrelated hypergraph with a Poisson degree distribution $P(q)$ and a shifted Poisson cardinality distribution $Q(m)$. Note that we need to shift the Poisson $Q(m)$ distribution since cardinalities $m=0,1$ are impossible: 
\bea
&&
P(q) = e^{-\avg{q}} \frac{\avg{q}^q}{q!}
,\nonumber \\
&&
Q(m\geq2) = e^{-(\avg{m}-2)} \frac{(\avg{m}-2)^{m-2}}{(m-2)!}
.
\label{Poisson}
\eea
Note that $\avg{m}$ and $
\avg{q}$ satisfy the equality $\avg{m}M=\avg{q}N$, so that $\avg{m}/\avg{q}=N/M$.
The generating functions for these distributions are 
\bea
&&
G_P(z) = e^{\avg{q}(z-1)} 
,
\nonumber 
\\[3pt]
&&
G_Q(z) = z^2 e^{(\avg{m}-2)(z-1)}
.
\label{44040}
\eea
%% 

%%%%%%%%%%%%%%%%%%%%%%%%%%%%%%%%%%%%%%%%%%%
%%%%%%%%%%%%%%%%%%%%%%%%%%%%%%%%%%%%%%%%%%%

\begin{figure}[t]
%%[htbp]
\begin{center}
\includegraphics[width=\columnwidth]{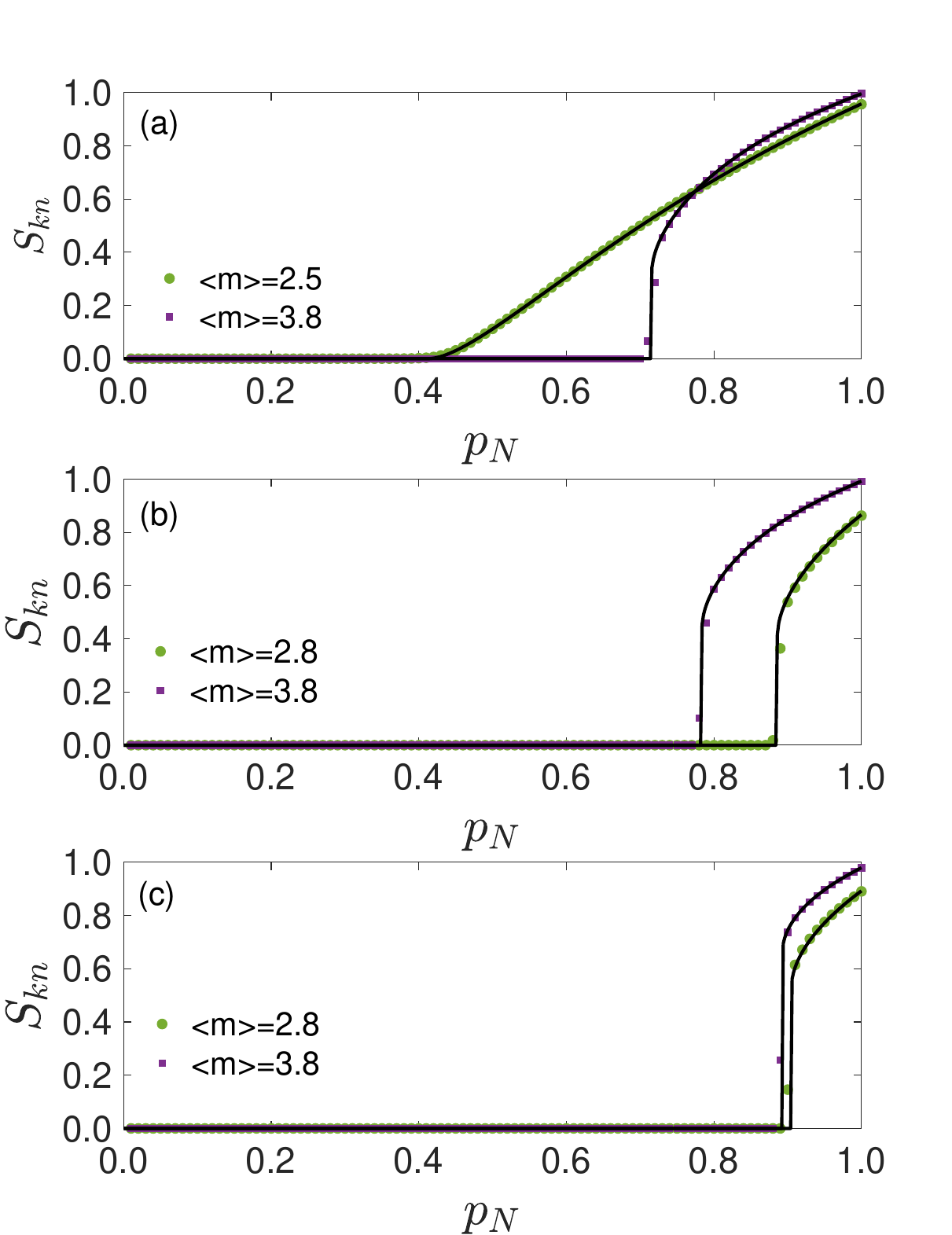}
\end{center}
\caption{Fraction of nodes in the $(k,n)$-core $S_{kn}$ as a function of $p_N$ for different $(k,n)$-cores and $p_H=1$. The $(k,n)$-core percolation is displayed for the: $(2,2)$-core (panel (a)), the $(2,3)$-core (panel (b)), and 
the $(3,2)$-core (panel (c)). The $(2,2)$ percolation transition is continuous for $\avg{m}=2.5$ and discontinuous for $\avg{m}=3.5$. Note that for $\avg{q}=2\avg{m}$ and $p_H=1$ the tricritical point of the $(2,2)$-core occurs at  $p_N=0.492143\ldots$, $\avg{m}=2.67731\ldots$. All hypergraphs have Poisson cardinality and degree distributions defined in Eq.~(\ref{Poisson}) with average degree $\avg{q}=2\avg{m}$ while $\avg{m}$ is indicated in the legend. Symbols indicate simulations obtained for  $N=10^4$ node hypergraphs averaged $100$ times, solid lines indicate our theoretical predictions.
}
\label{fsim}
\end{figure}

%%%%%%%%%%%%%%%%%%%%%%%%%%%%%%%%%%%%%%%%%%%
%%%%%%%%%%%%%%%%%%%%%%%%%%%%%%%%%%%%%%%%%%%

In Fig.~\ref{fsim} we show the sizes of the $(2,2)$, $(2,3)$, and $(3,2)$-cores obtained from MonteCarlo simulations on random hypergraphs with the Poisson cardinality and degree distributions given by Eq.~(\ref{Poisson}) with $\avg{q}=2\avg{m}$ and $p_H=1$. The simulations are in excellent agreement with our theoretical results and demonstrate that the $(2,2)$-core percolation can display both continuous and discontinuous transitions. Note that for $\avg{q}=2\avg{m}$ and $p_H=1$ the tricritical point of the $(2,2)$-core occurs at $p_N=0.492143\ldots$,  $\avg{m}=2.67731\ldots$. 

%%%%%%%%%%%%%%%%%%%%%%%%%%%%%%%%%%%%%%%%
%%%%%%%%%%%%%%%%%%%%%%%%%%%%%%%%%%%%%%%%

\begin{figure}[t]
%%[htbp]
\begin{center}
\includegraphics[width=\columnwidth]{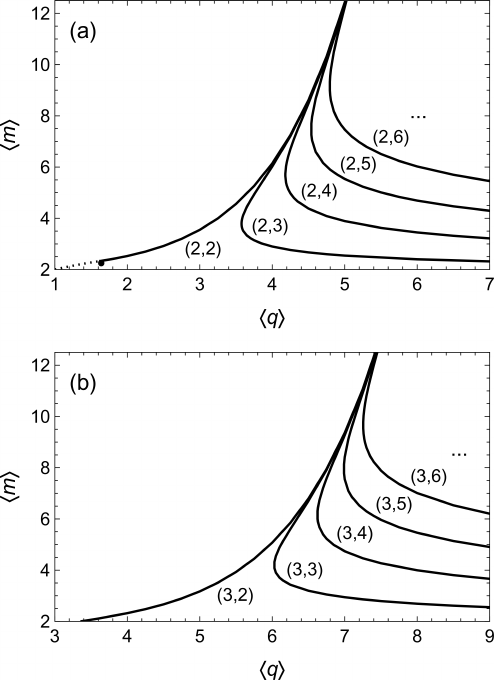} 
\end{center}
\caption{The phase diagram of $(k,n)$-core percolation in the $\avg{q}$--$\avg{m}$ plane is shown for  uncorrelated Poisson hypergraphs with $p_H=p_N=1$. Panel (a) displays the phase diagram for $(k,n)$-core percolation with $(k,n)$ given by  $(2,2)$, $(2,3)$, $(2,4)$, $(2,5)$, and $(2,6)$; panel (b) displays the core percolation phase diagram for $(k,n)$  given by   $(3,2)$, $(3,3)$, $(3,4)$, $(3,5)$, and $(3,6)$. 
 Each $(k,n)$-core exists in the whole region to the right of the corresponding boundary. 
All boundaries are discontinuous transitions with one exception, namely, the leftmost dotted piece of the boundary for the $(2,2)$-core, which is a continuous phase transition. The tricritical point at the $(2,2)$-core's phase boundary is $\avg{q}_\text{tricritical} = 1.628\ldots=7e^{1/3}/6$, $\avg{m}_\text{tricritical} = 2.333\ldots=7/3$.  
For the $(2,2)$-core, the phase boundary ends at the point $\avg{q}=1$, $\avg{m}=2$. 
For the $(3,2)$-core, the phase boundary ends at the point $\avg{q}=3.350919$, $\avg{m}=2$.
}
\label{fS}
\end{figure}

%%%%%%%%%%%%%%%%%%%%%%%%%%%%%%%%%%%%%%%%
%%%%%%%%%%%%%%%%%%%%%%%%%%%%%%%%%%%%%%%%

The phase diagram of the $(k,n)$-core percolation for  $p_H=p_N=1$, is shown in 
Fig.~\ref{fS} in the  $\avg{q}$--$\avg{m}$ phase space for $(2,n)$- and $(3,n)$-cores.  
The   difference from the $k$-core problem for ordinary graphs,  where the phase transition is continuous for $k=2$ and discontinuous for $k\geq3$ is apparent. 
Indeed, for random hypergraphs, the $(2,2)$-core phase boundary consists of 
%%percolation transition, we have 
two lines---of a continuous transition (dotted) and of a discontinuous one (solid)---converging at the tricritical point with the following coordinates: 
\bea
\avg{q}_\text{tricritical} &=& \frac{7}{6}\, e^{1/3} = 1.628\ldots 
, 
\nonumber
\\[3pt]
\avg{m}_\text{tricritical} &=& \frac{7}{3} = 2.333\ldots
.
\label{44050}
\eea
The phase boundary for $(2,n)$- and $(3,n)$-core percolation can be obtained by imposing that $\avg{m}=2$, which corresponds to the minimum possible value for the average cardinality of the hyperedges. Indeed for $\avg{m}=2$ the hypergraph reduces to an ordinary network. 

%%%%%%%%%%%%%%%%%%%%%%%%%%%%%%%%%%%%%%%%
%%%%%%%%%%%%%%%%%%%%%%%%%%%%%%%%%%%%%%%%

\begin{figure*}[t]
%%[htbp]
\begin{center}
\includegraphics[width=1.8\columnwidth]{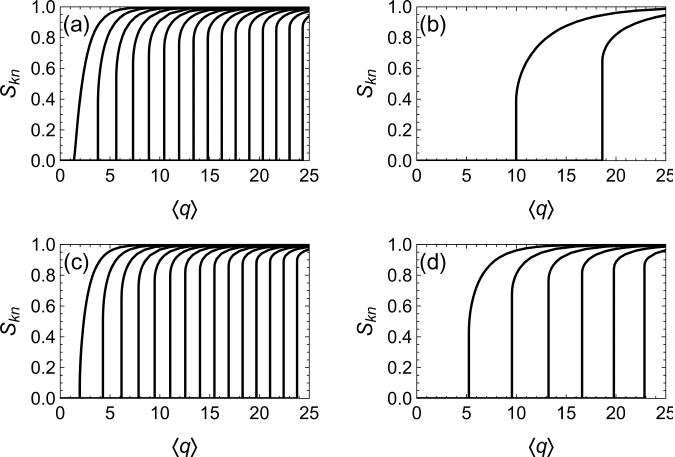} 
\end{center}
\caption{Relative sizes $S_{kn}$ of the $(k,n)$-cores vs. $\avg{q}$ for $\avg{m} = 2.2$ (panels (a) and (b)) and $\avg{m} = 2.5$ (panels (c) and (d)) plotted for $p_N=p_H=1$. 
Panel (a): the  curves from the left to the right display the $(k,n)$-core percolation for  $(k,n)$ given by  $(2,2)$, $(3, 2)$, $(4, 2)$, $(5, 2)$, $(6, 2)$, $(7, 2)$, $(8, 2)$, $(9, 2)$, $(10, 2)$, $(11, 2)$, $(12, 2)$, $(13, 2)$, $(14, 2)$, $(15, 2)$, $(16, 2)$, and $(17, 2)$; panel (b) the  curves from left to right display the $(k,n)$-core percolation for $(k,n)$  given by  $(2, 3)$ and $(3, 3)$. 
The phase transition for the $(2,2)$-core for $\avg{m} = 2.2$ is continuous, and for the other $(k,n)$-cores the transitions are hybrid. 
Panel (c): the  curves from the left to the right display the $(k,n)$-core percolation with  $(k,n)$ given by  $(2,2)$, $(3, 2)$, $(4, 2)$, $(5, 2)$, $(6, 2)$, $(7, 2)$, $(8, 2)$, $(9, 2)$, $(10, 2)$, $(11, 2)$, $(12, 2)$, $(13, 2)$, $(14, 2)$, $(15, 2)$, and $(16, 2)$; panel (d): the  curves from left to right display $(k,n)$-core percolation with $(k,n)$ given by  $(2, 3)$,  $(3, 3)$, $(4, 3)$, $(5, 3)$, $(6, 3)$, and $(7, 3)$. 
}
\label{f4}
\end{figure*}

%%%%%%%%%%%%%%%%%%%%%%%%%%%%%%%%%%%%%%%%
%%%%%%%%%%%%%%%%%%%%%%%%%%%%%%%%%%%%%%%%

For the $(2,2)$-core, the phase boundary ends at the point $\avg{q}=1$, $\avg{m}=2$. 
 {One can see in Fig.~\ref{f3}(b) that for the $(3,2)$-core, the phase boundary ends at a point on a line $\avg{m}=2$. The coordinate $\avg{q}$ of this point can be obtained exactly. 
We substitute the degree and cardinality distributions, Eq.~\eq{Poisson}, into the equations Eq.~\eq{440000} for $k=3$, $n=2$, and $\avg{m}=2$ [one can conveniently use the generating functions of the distributions, Eq.~\eq{44040}]. This results in the following equations: 
\bea
&&
%%\left( \avg{q} + 1 - \sqrt{(\avg{q} - 1)^2 -4} \right)
V^\star = 1 - (1 + \avg{q}V^\star) e^{-\avg{q}V^\star}
,
\nonumber
\\[3pt]
&&
1 = \avg{q}^2 V^\star e^{-\avg{q}V^\star}
%%.
\label{44055}
\eea
for $\avg{q}$ and $V^\star$. 
Excluding $V^\star$ from Eq.~\eq{44055}, 
\be
\avg{q}V^\star = \frac{1}{2} \left[ \avg{q} - 1 + \sqrt{(\avg{q} - 1)^2 -4} \right]
,
\label{44057}
\ee
%%
%%Excluding $V^\star$ from Eq.~\eq{44055}, 
we get the equation for $\avg{q}$:  
%%
%%For the $(3,2)$-core, the phase boundary ends at the point $\avg{q}=3.350919$, $\avg{m}=2$, where this value of $\avg{q}$ is the root of the equation: 
%%
\bea
&&
%%\left( \avg{q} + 1 - \sqrt{(\avg{q} - 1)^2 -4} \right)
e^{\avg{q} - 1 + \sqrt{(\avg{q} - 1)^2 -4}}
\nonumber
\\[3pt]
&&
= \avg{q}^2 \frac{\avg{q} - 1 + \sqrt{(\avg{q} - 1)^2 -4}}{\avg{q} - 1 - \sqrt{(\avg{q} - 1)^2 -4}}
,
\label{44060}
\eea
whose root $\avg{q}=3.350919$. 
Thus the phase boundary of the $(3,2)$-core ends at the point $\avg{q}=3.350919$, $\avg{m}=2$.
}

Figure~\ref{f4} shows the dependencies of the relative sizes $S_{kn}$ of the $(k,n)$-cores on $\avg{q}$ for different values of mean cardinality $\avg{m}$ for the Poisson hypergraph with $p_N=p_H=1$.

%%%%%%%%%%%%%%%%%%%%%%%%
%%%%%%%%%%%%%%%%%%%%%%%%
%%%%%%%%%%%%%%%%%%%%%%%%

\section{Hypergraphs $(k,n)$-core second-neighbor problems  and their pruning algorithm}
\label{second_neighbor}

\subsection{The second-neighbor pruning algorithm}

%Until now we have defined the $(k,n)$-core of a the hypergraph based on a pruning algorithm that prunes nodes considering the connectivity of the hyperedges they belong to and  prunes hyperedges according to the connectivity of the nodes belonging to it.
Until now we have defined the $(k,n)$-core of a the hypergraph based on a pruning algorithm that prunes nodes and   hyperedges according to their connectivity.
However there is another possibility, i.e. pruning either nodes of hyperedges depending on the state of  their  second neighbors in the factor graph.
This implies a new set of algorithms pruning  nodes considering the connectivity of the hyperedges they belong to or pruning  hyperedges according to the connectivity of the nodes belonging to it.

To this end we distinguish two types of second-neighbor $(k,n)$-core problems: in the first nodes are iteratively pruned and in the second hyperedges are iteratively pruned.
We note that there is no symmetry between these two pruning algorithms.
This is due to the fact that in   hypergraph percolation hyperedges  in order to be belong to the giant component must have  all their nodes  undamaged while no corresponding constraint holds for the nodes. 
Interestingly we will observe important differences between the second-neighbor $(k,n)$-cores obtained pruning only nodes and the ones obtained pruning only hyperedges.
 
Let us consider these two algorithms and their corresponding message-passing equations separately.

\subsection{The message-passing equations for the second-neighbor node pruning algorithm}

Let us start from a configuration in which we  initially damage  nodes with probability $1-p_N$ and/or hyperedges with probability $1-p_H$.
If we consider the pruning on the nodes the second-neighbor hypergraph $(k,n)$-core can be  obtained using the following  pruning algorithm:
\begin{itemize}
\item[(1)] 
Damage iteratively all nodes belonging to less than $k$  hyperedges each connected to at least   $n$ (undamaged) nodes.
\item[(2)] 
Define the $(k,n)$-core as the giant component of the network induced by the undamaged nodes and their connected hyperedges.
\end{itemize}
As we will see in the following this algorithm is 
very 
closely related to the algorithm defined in Sec.~\ref{first_neighbor}. Note that also in this case, as in the algorithm defined in Sec.~\ref{first_neighbor}, due to the definition of hypergraph giant component, every hyperedge of cardinality $m\geq n$ belonging to the $(k,n)$-core will be connected to exactly $m\geq n$ undamaged nodes. Therefore the result of the algorithm is unchanged if only hyperedges of cardinality less than $n$ are pruned at stage (1).

In order to derive the corresponding message-passing equation we start with the message-passing equations \cite{bianconi2023theory} implementing hypergraph percolation at step (2).
Using the same notation used in Sec.~\ref{first_neighbor} we see therefore that the messages $\hat{w}_{i\rightarrow \alpha},\hat{v}_{\alpha\to i}$ obey 
\bea
\hat{w}_{i\rightarrow \alpha} &=s_i&  \left[ 1 - \prod_{\beta\in N(i)\setminus \alpha}(1 - \hat{v}_{\beta\rightarrow i}) \right]
,
\label{hwv} 
\\
\hat{v}_{\alpha\to i} &=&y_{\alpha}\left(\prod_{j\in N(\alpha)\setminus i}\!s_j\!\right)\left[ 1 -\prod_{j\in N(\alpha)\setminus i}\!(1 - \hat{w}_{j\rightarrow \alpha}) \right],
\nonumber
\eea
where $s_i$ indicates whether node $i$ has been damaged/pruned, $s_i=0$, or not, $s_i=1$.
The indicator functions $\hat\sigma_i\in \{0,1\}$ and $\hat{r}_{\alpha}\in \{0,1\}$, indicating whether nodes and hyperedges are in the giant component and hence in the $(k,n)$-core, are expressed in terms of $s_i$, $x_i$, and $y_{\alpha}$, are  given respectively by the equations 
\bea
\hat{\sigma}_i&=&s_i \left[ 1 - \prod_{\beta\in N(i)}(1 - \hat{v}_{\beta\rightarrow i}) \right]
,
\nonumber 
\\
\hat{r}_{\alpha} &=&y_{\alpha}\left(\prod_{j\in N(\alpha)}\!s_j\!\right)\left[ 1 -\prod_{j\in N(\alpha)}\!(1 - \hat{w}_{j\rightarrow \alpha}) \right]
.
\eea 
The pruning of the nodes determines the indicator function $s_i$ which is  only non-zero if the node $i$ receives at least $k$ positive messages from its hyperedge neighbors of cardinality at least $n$, i.e. 
\bea
s_i=x_i\sum_{\substack{\Theta \subseteq N(i) \\ |\Theta| \geq k}}\ \ \prod_{\beta \in \Theta} \tilde{v}_{\beta \to i}\prod_{\gamma \in N(i)\setminus \Theta} (1 - \tilde{v}_{\gamma \to i})
,
\label{si_np}
\eea
where we have defined  
\bea
\tilde{v}_{\alpha\to i}=\theta(|\alpha|-n)\hat{v}_{\alpha\to i} 
\eea
with $\theta(x)=1$ if $x\geq 0$ and $\theta(x)=0$ otherwise.
Inserting this expressions of $s_i$  into Eq.~(\ref{hwv}), and taking into account that the messages $\hat{v}_{\alpha\to i}$ are defined under the assumption that node $i$ is in the giant hypergraph component and the messages $\hat{w}_{i\to \alpha}$ are defined under the assumption that hyperedge $\alpha$ is in the giant component,    exploiting the fact that the messages take only $0$, $1$ values we get
\bea
\hat{w}_{i\to \alpha}&=&x_i\sum_{\substack{\Theta \subseteq N(i)\setminus\alpha \\ |\Theta| \geq k-1}}\ \ \prod_{\beta \in \Theta} \tilde{v}_{\beta \to i}\prod_{\gamma \in N(i)\setminus (\Theta \cup \alpha)} (1 - \tilde{v}_{\gamma \to i})
,
\nonumber \\
\tilde{v}_{\alpha\to i} &=&y_{\alpha}\theta(|\alpha|-n) \!\! \prod_{j\in N(\alpha)\setminus i}\!\hat{w}_{j\rightarrow \alpha}
. 
\label{Hhv}
\eea

Exploiting furthermore the fact that both $s_i$ and the messages $\hat{w}_{i\to \alpha},\hat{v}_{\alpha\to i}$ take values $0$, $1$, it is also immediate to show that 
\bea
\hat{\sigma}_i&=&s_i=x_i\sum_{\substack{\Theta \subseteq N(i) \\ |\Theta| \geq k}}\ \ \prod_{\beta \in \Theta} \tilde{v}_{\beta \to i}\prod_{\gamma \in N(i)\setminus \Theta} (1 - \tilde{v}_{\gamma \to i}),\nonumber \\
\hat{r}_{\alpha}&=&y_{\alpha}\prod_{j\in N(\alpha)}\!\hat{w}_{j\rightarrow \alpha}. 
\eea

\begin{figure}[t]
\begin{center}
\includegraphics[width=1\columnwidth]{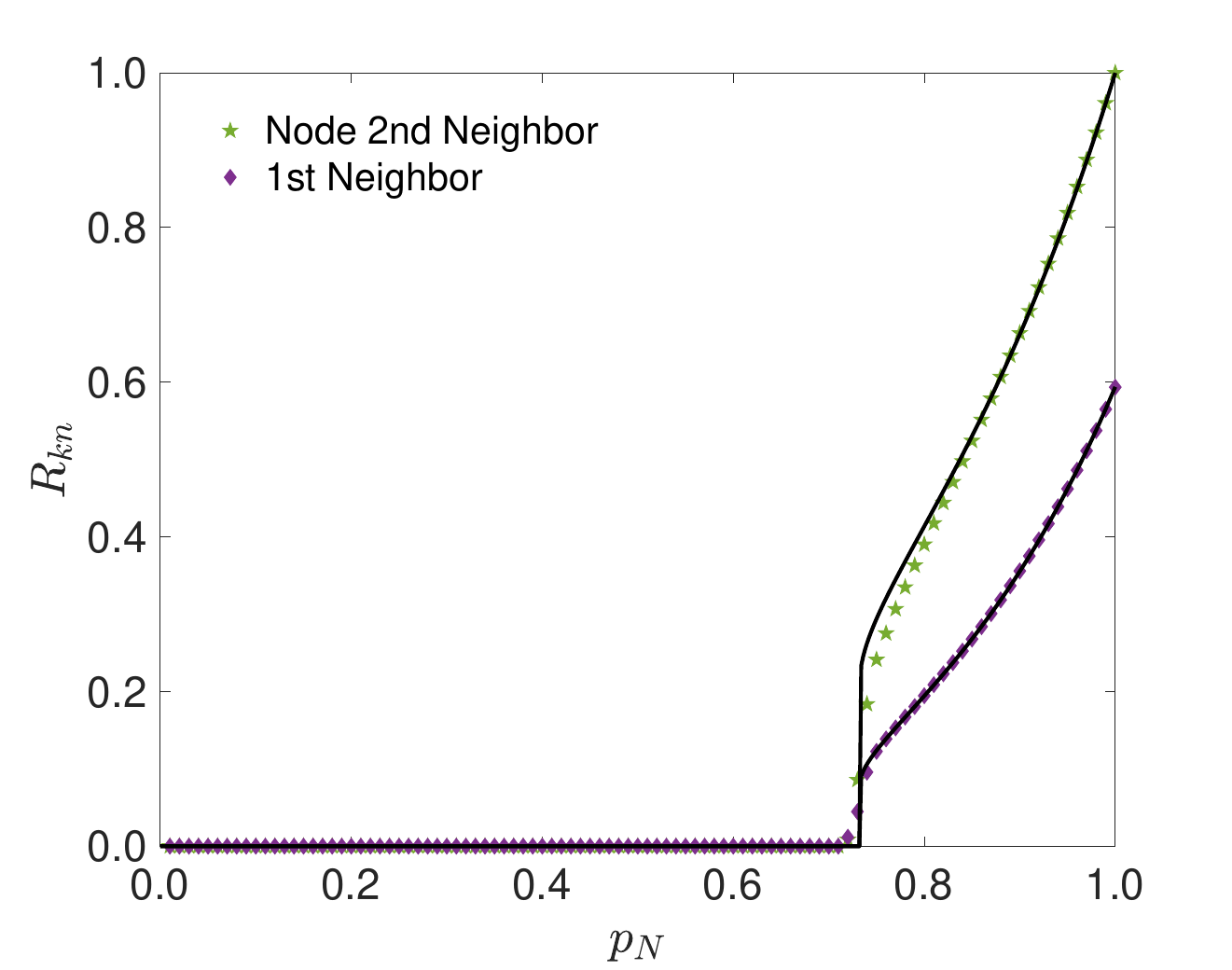}
\end{center}
\caption{The fraction of hyperedges $R_{kn}$ in the $(2,4)$-core is plotted versus $p_N$ for the 2nd neighbor node $(2,4)$-core  and for the 1st neighbor $(2,4)$-core percolation. The percolation threshold is the same but $R_{kn}$ differ. The hypergraph have Poisson cardinality and degree distribution with $\avg{m}=4$, $\avg{q}=16$ and number of nodes $N=5000$. The symbols correspond to Monte Carlo simulations averaged over $100$ iterations for $p_H=1$. The solid lines are the theoretical predictions.}
\label{f_Rkn}
\end{figure}

Another set of message-passing equations holds when we do not know direct access to the configuration of the initial damage $\{x_i\}, \{y_{\alpha}\}$ but we only know the probability that nodes and hyperedges are initially intact, i.e. $p_N$ and $p_H$ respectively. This second set of message-passing equations can be simply obtained by averaging the messages over the  initial damage distribution 
$P(\{x_i\},\{y_{\alpha}\})$ given by Eq.~(\ref{random_xy}).
In this way we  obtain the following set of message-passing equations (note that the messages $w_{i\to\alpha}$ and $v_{\alpha\to i}$ now take real values between $0$ and $1$): 
\bea
{w}_{i\to \alpha}&=&p_N\sum_{\substack{\Theta \subseteq N(i)\setminus\alpha \\ |\Theta| \geq k-1}}\ \ \prod_{\beta \in \Theta} {v}_{\beta \to i}\prod_{\gamma \in N(i)\setminus (\Theta\cup \alpha)} (1 - {v}_{\gamma \to i})
,
\nonumber 
\\
{v}_{\alpha\to i} &=&p_H\theta(|\alpha|-n)\prod_{j\in N(\alpha)\setminus i}\!{w}_{j\rightarrow \alpha}
. 
\label{Hp_hv}
\eea
The probability $\sigma_i$ that the node $i$  belongs to the $(k,n)$-core  and the probability $r_{\alpha}$ that the hyperedge $\alpha$ belongs to the $(k,n)$-core are given by
\bea
{\sigma_i}&=&p_N\sum_{\substack{\Theta \subseteq N(i) \\ |\Theta| \geq k}}\ \ \prod_{\beta \in \Theta} {v}_{\beta \to i}\prod_{\gamma \in N(i)\setminus \Theta} (1 - {v}_{\gamma \to i}),
\nonumber 
\\
{r_{\alpha}}&=&p_H\prod_{j\in N(\alpha)}\!{w}_{j\rightarrow \alpha}
.
\eea
The fraction $S_{kn}$ of nodes in the $(k,n)$-core is given by 
\bea
S_{kn}=\frac{1}{N}\sum_{i=1}^N\sigma_i
\label{MSkn}
\eea
and the fraction $R_{kn}$ of hyperedges in the $(k,n)$-core is given by 
\bea
R_{kn}=\frac{1}{M}\sum_{\alpha=1}^Mr_{\alpha}
.
\label{MRkn}
\eea
Therefore this algorithm essentially reduced to the first-neighbor $(k,n)$-core studied in Sec.~\ref{first_neighbor}. Indeed the percolation threshold and the nature of the transition is the same, the fractions of nodes within these $(k,n)$-cores, $S_{kn}$, also coincide while only  the fractions of hyperedges within the $(k,n)$-cores, $R_{kn}$, can in general differ, see Fig.~\ref{f_Rkn}. 

%%%%%%%%%%%%%%%%%%%%%%%%%%%%%%%%%%%%%%%%%%%%%
%%%%%%%%%%%%%%%%%%%%%%%%%%%%%%%%%%%%%%%%%%%%%

%%%%%%%%%%%%%%%%%%%%%%%%%%%%%%%%%%%%%%%%%%%%%
%%%%%%%%%%%%%%%%%%%%%%%%%%%%%%%%%%%%%%%%%%%%%

\subsection{The message-passing equations for the second-neighbor hyperedge pruning algorithm}

We start 
%%always 
from the configuration in which we initially damage either nodes with probability $1-p_N$ and/or hyperedges with probability $1-p_H$.
If we consider the pruning on the hyperedges of  the hypergraph, the $(k,n)$-core can be  obtained  using the following  pruning algorithm: 
\begin{itemize}

\item[(1)] 
Damage iteratively all hyperedges having less than $n$ nodes each connected to at least $k$ undamaged hyperedges.

\item[(2)] 
Define the $(k,n)$-core as the giant component of the network induced by the undamaged hyperedges.

\end{itemize}
This $(k,n)$-core is the maximal connected subhypergraph each of whose hyperedges has at least $n$ nodes with degrees at least $k$. Note that this pruning algorithm doesn't change the number of nodes in the network. It stays equal to $N$. 

Our starting point is always the set of message-passing equations for hypergraph percolation \cite{bianconi2023theory} where now at step (2) of the pruning process each hyperedge $\alpha$ is either damaged ($s_{\alpha}=0$) or not damaged ($s_{\alpha}=1$).
Using always the same notations we have been using so far we get:
\bea
\hat{w}_{i\rightarrow \alpha} &=x_i&  \left[ 1 - \prod_{\beta\in N(i)\setminus \alpha}(1 - \hat{v}_{\beta\rightarrow i}) \right]
,
\label{hvw_H} 
\\
\hat{v}_{\alpha\to i} &=&s_{\alpha}\left(\prod_{j\in N(\alpha)\setminus i}\!x_j\!\right)\left[ 1 -\prod_{j\in N(\alpha)\setminus i}\!(1 - \hat{w}_{j\rightarrow \alpha}) \right]
.
\nonumber 
\eea
The indicator functions $\hat\sigma_i\in \{0,1\}$ and $\hat{r}_{\alpha}\in \{0,1\}$ indicating whether nodes and hyperedges are in the giant component at  step (2) and hence in the $(k,n)$-core are expressed in terms of $x_i$ and $s_{\alpha}$,  and they are given, respectively, by the equations 
\bea
\hat{\sigma}_i&=&x_i \left[ 1 - \prod_{\beta\in N(i)}(1 - \hat{v}_{\beta\rightarrow i}) \right],
\nonumber 
\\
\hat{r}_{\alpha} &=&s_{\alpha}\left(\prod_{j\in N(\alpha)}\!x_j\!\right)\left[ 1 -\prod_{j\in N(\alpha)}\!(1 - \hat{w}_{j\rightarrow \alpha}) \right]
.
\eea 
The pruning of the hyperedge determines the indicator function $s_\alpha$. The indicator function $s_\alpha$ is  only non-zero if the hyperedge is connected to al least $n$ nodes, each connected to at least $k$ undamaged hyperedges (belonging to  the $(k,n)$-core/giant component). Let us define a node $i$ to be active if it receives at least $k$ positive messages from its neighbors, i.e. 
\bea
a_i=x_i\sum_{\substack{\Theta \subseteq N(i) \\ |\Theta| \geq k}} \prod_{\beta \in \Theta} \hat{v}_{\beta \to i}\prod_{\gamma \in N(i)\setminus \Theta} (1 - \hat{v}_{\gamma \to i})
.
\label{ai}
\eea
The damage of the hyperedges is therefore determined by the activity of their nodes by 
\bea
\hspace{-8mm}s_{\alpha}=y_\alpha\!\!\sum_{\substack{\Theta \subseteq N(\alpha) \\ |\Theta| \geq n}} \prod_{r \in \Theta} a_r\hat{w}_{r \to \alpha}\prod_{ r\in N(\alpha)\setminus \Theta} (1 - a_r\hat{w}_{r \to \alpha})
.
\label{010}
\eea
Let us define $\tilde{w}_{i\to \alpha}$ as 
\bea
\tilde{w}_{i\to\alpha}=a_i\hat{w}_{i\to\alpha}
.
\eea
In order to express $\hat{v}_{\alpha\to i}$ we need to distinguish the case in which $i$ is active, i.e. $a_i=1$ and the case in which $a_i$ is not active.
Moreover, using the fact that $s_{\alpha}, a_i$ and the message all take values $0$, $1$ we obtain that 
\bea
&&
\hat{v}_{\alpha\to i}=\tilde{y}_{\alpha,i}a_i \!\! 
\sum_{\substack{\Theta \subseteq N(\alpha)\setminus i \\ |\Theta| \geq n-1}} \ \prod_{r \in \Theta} \tilde{w}_{r \to \alpha}\!\prod_{ r\in N(\alpha)\setminus (\Theta\cup i)}\! (1 - \tilde{w}_{r \to \alpha})
\nonumber 
\\
&&
+\tilde{y}_{\alpha,i}(1-a_i) \!\!
\sum_{\substack{\Theta \subseteq N(\alpha)\setminus i \\ |\Theta| \geq n}} \ \prod_{r \in \Theta} \tilde{w}_{r \to \alpha}\!\prod_{ r\in N(\alpha)\setminus (\Theta\cup i)}\! (1 - \tilde{w}_{r \to \alpha})
,
\nonumber 
\eea
where $\tilde{y}_{\alpha,i}$ is given by 
\bea
\tilde{y}_{\alpha,i}=y_\alpha\left(\prod_{j\in N(\alpha)\setminus i}\!x_j\!\right)
\eea
and $\tilde{w}_{i\to \alpha}$ is given by 
\bea
\hspace{-18pt}\tilde{w}_{i\to \alpha}=a_i \!\! \sum_{\substack{\Theta \subseteq N(i)\setminus \alpha \\ |\Theta| \geq k-1}}\ \prod_{\beta \in \Theta} \hat{v}_{\beta \to i}\!\prod_{\gamma \in N(i)\setminus (\Theta\cup \alpha)}\! (1 - \hat{v}_{\gamma \to i}).
\eea
Similarly one can show that the indicator functions $\hat{\sigma}_i$ and $\hat{r}_{\alpha}$ are given by 
\bea
\!\!\!\!\!\!\!\!
\hat{\sigma}_i&=&x_i\left[1-\prod_{\alpha\in N(i)}\hat{v}_{\alpha\to i}\right]
,
\nonumber 
\\
\!\!\!\!\!\!\!\!
\hat{r}_{\alpha}&=&{\hat{y}}_{\alpha}\sum_{\substack{\Theta \subseteq N(\alpha) \\ |\Theta| \geq n}} \ \prod_{r \in \Theta} \tilde{w}_{r \to \alpha}\prod_{r \in N(i)\setminus \Theta} (1 - \tilde{w}_{r \to \alpha})
, 
\eea
where 
\bea
{\hat{y}}_{\alpha}=y_{\alpha}\left(\prod_{j\in N(\alpha)} x_j\right)
.
\eea

We now derive the second set of  message-passing equations that hold when we do not have  direct access to the configuration of the initial damage $\{x_i\}, \{y_{\alpha}\}$ but we only know the probability that nodes and hyperedges are initially intact, i.e. $p_N$ and $p_H$ respectively by averaging the messages over the  initial damage distribution 
$P(\{x_i\},\{y_{\alpha}\})$. First of all we observe that $\tilde{w}_{i\to\alpha}$ are only non-zero if the node $i$ is active. Therefore we consider only the average  message $v_{\alpha\to i}=\Avg{a_i\hat{v}_{\alpha\to i}}$ and the average message $w_{i\to \alpha}=\Avg{\tilde{w}_{i\to \alpha}}$  which constitute the closed form equations determining the percolation threshold.
In this way, paying attention to the fact that the messages take real values between $0$ and $1$, we  obtain the following set of message-passing equations:
\bea
\hspace{-10mm}
{v}_{\alpha\to i} &=&p_{HN}\!\sum_{\substack{\Theta \subseteq N(\alpha)\setminus i \\ |\Theta| \geq n-1}}\ \prod_{r \in \Theta} {w}_{r \to \alpha}\!\!\prod_{r \in N(\alpha)\setminus (\Theta\cup i)}\!\! (1 - {w}_{r \to \alpha})
,
\nonumber 
\\[3pt]
{w}_{i\to \alpha}&=&\sum_{\substack{\Theta \subseteq N(i)\setminus\alpha \\ |\Theta| \geq k-1}}\ \prod_{\beta \in \Theta} {v}_{\beta \to i}\!\!\prod_{\gamma \in N(i)\setminus (\Theta\cup\alpha)} \!\!(1 - {v}_{\gamma \to i})
,
\label{H2p__hw}
\eea
where $p_{HN}=p_H p_N^{m-1}$.

The probability  $r_{\alpha}$ that the  hyperedge $\alpha$ belongs to the $(k,n)$-core is given by
\bea
\hspace{-4mm}{r_{\alpha}}&=&p_{HN}\sum_{\substack{\Theta \subseteq N(\alpha) \\ |\Theta| \geq n}}\ \prod_{r \in \Theta} {w}_{r \to \alpha}\prod_{r \in N(\alpha)\setminus \Theta} (1 - {w}_{r \to \alpha}). 
\label{10000}
\eea
However we need some additional care to express the probability $\sigma_i$ that node $i$ belongs to the $(k,n)$-core. In particular,  the giant component will include all active nodes and the  inactive nodes that are intact and are connected to at least one undamaged hyperedge. Note that an hyperedge including  an inactive node can only be active if at least $n$ of its other nodes are active. This implies that 
the all intact  nodes will belong to the giant component unless both  (i) and (ii) are satisfied. These two conditions are: (i) the node is not connected to any hyperedge including at least to $n$ other active nodes; (ii) the nodes belongs to less than  $k$ hyperedges linked to  $n-1$ other active nodes. It follows from this that 
the probability that a node belong to the $(k,n)$-core is 
\bea
&&
\hspace{-19pt}\sigma_i=p_N
\nonumber 
\\[3pt]
&&
\hspace{-14pt}
\times \left[1-\left(\sum_{\substack{\Theta \subseteq N(i) \\ |\Theta| \leq k-1}}\ \prod_{r \in \Theta} {\theta}_{r \to i}\prod_{ r\in N(\alpha)\setminus \Theta} (1 -{\rho}_{\alpha \to i})\right)\right]
\!\!, 
\label{10001}
\eea
where $\theta_{\alpha\to i}$ and $\rho_{\alpha\to i}$ are given by 
\bea
\theta_{\alpha\to i} &=& p_{HN} \!\! 
\sum_{\substack{\Theta \subseteq N(\alpha)\setminus i \\ |\Theta|= n-1}}\ \prod_{r \in \Theta} {w}_{r \to \alpha} \!\! \prod_{ r\in N(\alpha)\setminus (\Theta\cup i)} \!\! (1 - {w}_{r \to \alpha})
,
\nonumber 
\\[3pt]
\rho_{\alpha\to i} &=&{\theta}_{\alpha \to i} 
\nonumber 
\\[3pt]
&&
\hspace{-25pt}
+p_{HN} \!\! 
\sum_{\substack{\Theta \subseteq N(\alpha)\setminus i \\ |\Theta| \geq n}}\ \prod_{r \in \Theta} {w}_{r \to \alpha} \!\! \prod_{ r\in N(\alpha)\setminus (\Theta\cup i)} \!\! (1 - {w}_{r \to \alpha})
.
%%\nonumber
%\\
%&&
\label{73000}
\eea
The fraction $S_{kn}$ of nodes in this $(k,n)$-core is given by 
\bea
S_{kn}=\frac{1}{N}\sum_{i=1}^N\sigma_i
\label{MSkn}
\eea
and the fraction $R_{kn}$ of hyperedges in the $(k,n)$-core is given by 
\bea
R_{kn}=\frac{1}{M}\sum_{\alpha=1}^Mr_{\alpha}
.
\label{MRkn}
\eea

\subsection{Discussion of differences between the first-neighbor and the second-neighbor pruning algorithm}

When we consider the second-neighbor pruning algorithms node pruning and hyperedges pruning give rise to very different definition of $(k,n)$-cores.

For the   random hypergraphs belonging to the  configuration model ensembles, the equations determining the average messages $W=\avg{w_{i\to\alpha}}$ and $V=\avg{v_{\alpha\to i}}$ of the second-neighbor $(k,n)$-core with pruning of 
%%the 
nodes are:
\bea
&&
\!\!\!\!\!\!\!\!\!\!\!\!
V = p_H \sum_{m\geq n}\frac{mQ(m)}{\avg{m}} W^{m-1} 
,
\nonumber
\\[3pt]
&&
\!\!\!\!\!\!\!\!\!\!\!\!
W =  p_N\sum_{q=k}^\infty \frac{q P(q)}{\avg{q}} \sum_{s=k-1}^{q-1} {q-1 \choose s} V^s (1 - V)^{q-1-s}
,
\label{700}
\eea
These are the same equations determining the average messages of the first-neighbor $(k,n)$-core algorithm. It follows that the phase diagram of the first-neighbor $(k,n)$-core percolation coincides with the phase diagram for the second-neighbor $(k,n)$-core percolation with pruning of the nodes.
However the order parameters might differ. Indeed the order parameter $S_{kn}=\avg{\sigma_i}$ and $R_{kn}=\avg{r_{\alpha}}$, where $\sigma_i$ and $r_{\alpha}$ are given by Eqs.~\eq{10001} and \eq{10000}, are:
\bea
&&
R_{kn} = p_H\sum_{m} Q(m) W^m  
,
\label{710b}
\\[3pt]
&&
S_{kn} =  p_N \sum_{q=k}^\infty P(q) \sum_{s=k}^q {q \choose s} V^s (1 - V)^{q-s}
.
\label{710}
\eea
Note that only Eq.~(\ref{710b}) differs from the corresponding equation determining the first-neighbor $(k,n)$-core percolation, Eq.~\eq{455}, while Eqs.~\eq{710} and \eq{460} coincide. Indeed, Eq.~(\ref{710b}) includes a sum extended to  hyperedges of arbitrary cardinality $m$ while in Eq.~\eq{455} the sum is extended only to hyperedges of cardinality $m\geq n$.
It follows that the order parameter $S_{kn}$ is unchanged if one consider first-neighbor $(k,n)$-core percolation or second-neighbor $(k,n)$-core percolation with node pruning, but the order parameter $R_{kn}$ can change for $n\geq 3$. In order to demonstrate this,  we show in Fig.~\ref{f_Rkn} Monte Carlo results for first-neighbor  $(k,n)$-core percolation and for second-neighbor $(k,n)$-core percolation with node pruning. The results are in very good agreement with our theoretical predictions.

The critical behavior in second-neighbor $(k,n)$-core percolation with pruning of the hyperedges is distinct from these results.  
%%In sharp contrast with these results, if one consider second-neighbor $(k,n)$-core percolation with pruning of the hyperedges the obtained transition has a very distinct critical behavior. 
Indeed not only the order parameters can differ from the first-neighbor $(k,n)$-core transition, but also the nature of the transition and its critical points.
Indeed, the system of equation determining the nature of the phase transition reads in this case as
\bea
&&
V {=} p_H \! \sum_{m\geq n} p_N^{m-1}\frac{mQ(m)}{\avg{m}} \! \sum_{s=n-1}^{m-1} \!{m{-}1 \choose s} W^s (1 {-} W)^{m-1-s}
,
\nonumber
\\[3pt]
&&
W =\sum_{q\geq k} \frac{q P(q)}{\avg{q}} \sum_{s=k-1}^{q-1} {q-1 \choose s} V^s (1 - V)^{q-1-s}
.
\label{720}
\eea
where $V$ and $W$ are the average messages.
The equation
determining the fraction of hyperedges, $R_{kn}$, in the second-neighbor $(k,n)$-core with pruning of hyperedges is given by
\bea
\!\!\!\!\!\! \!\!\!\!\!\! 
&&
R_{kn} = p_H \! \sum_{m\geq n} \! p_N^{m}\,Q(m)  \sum_{s=n}^{m} {m \choose s} W^s (1 - W)^{m-s}
.
\label{725}
\eea
Moreover, the equation determining the fraction of nodes $S_{kn}$ in the second-neighbor $(k,n)$-core with pruning of hyperedges is more subtle.
These equations are: 
\bea
&&
\hspace{-19pt}
S_{kn} =  p_N 
\nonumber  
\\[3pt]
&&
\hspace{-19pt}
\times\sum_{q} P(q) \!\!\left[1- \!\!\!\!\sum_{s\leq \min(k-1,q)} {q \choose s} \tilde{V}^s (1 - \tilde{V}-\hat{V})^{m-s}  \right]
\!
,
\label{730}
\eea
where $\hat{V}$ and $\tilde{V}$  
are given by 
\bea
&&
\hat{V} = p_H \!\!\!\!\! \sum_{m\geq n+1} \!\!\!\! p_N^{m-1}\frac{mQ(m)}{\avg{m}}  \sum_{s=n}^{m-1} \! {m{-}1 \choose s} W^s (1 {-} W)^{m-1-s}
\!
,
\nonumber
\\[3pt]
&&
\tilde{V} = p_H \! \sum_{m\geq n} p_N^{m-1}\frac{mQ(m)}{\avg{m}}   {m-1 \choose n-1} W^{n-1} (1 - W)^{m-s}
.
\nonumber
\\
&&
\label{7301}
\eea
The rationale behind Eqs.~\eq{730} and \eq{7301} was explained while deriving the message-passing Eqs.~\eq{10001} and \eq{73000}, from which these equations directly follow. 

Intuitively an active node will be always part of the 2nd neighbor node $(k,n)$ core. An intact inactive node will be part of the $(k,n)$ core  only if it belongs to at least one hyperedge that belongs to the $(k,n)$ core. It follows that a node will be always in 2nd neighbor node $(k,n)$ core unless (i) none of its hyperedges is connected to at least $n$ other active nodes- which occurs with probability $\hat{V}$- and (ii) there are less than $k$ hyperedges connected to $n-1$ active nodes-which occurs with probability $\tilde{V}$. Note that condition (ii) together with (i) ensures that the node is not active.

\begin{figure}[t]
%%[htbp]
\begin{center}
\includegraphics[width=\columnwidth]{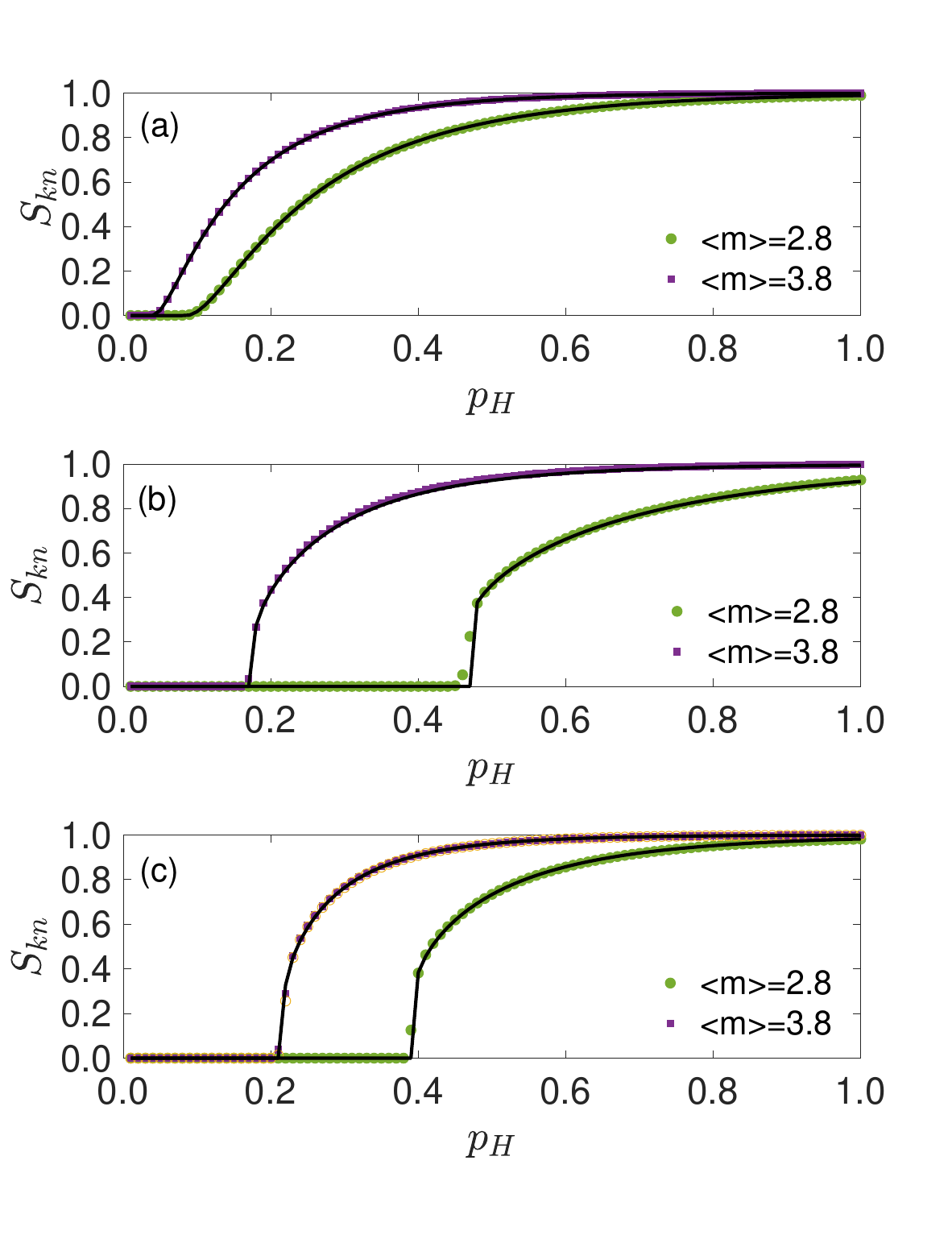} 
\end{center}
\caption{The fraction $S_{kn}$ of nodes in the 2nd neighbor $(k,n)$-core with hyperedge pruning is plotted versus $p_H$ for different values of $k$ and $n$: $(k,n)=(2,2)$ (panel (a)); $(k,n)=(2,3)$ (panel (b)); $(k,n)=(3,2)$ (panel (c). The hypergraphs have Poisson cardinality and degree distributions given by Eq.~(\ref{Poisson}), $\avg{q}=2\avg{m}$, $\avg{m}$ indicated in the legend, and $N=10^4$ nodes. }
\label{f_2nd_hyperedge}
\end{figure}

The phase diagram of 2nd neighbor $(k,n)$-core percolation with pruning of the hyperdeges is very different from the phase diagram for 2nd neighbor $(k,n)$-core percolation with pruning of the nodes. In particular, the phase transition is continuous if and only if $(k,n)=(2,2)$ with the second order phase transition line obtained for 
\bea
1=p_H\frac{\Avg{m(m-1)p_N^{m-1}}}{\Avg{m}}\frac{\Avg{q(q-1)}}{\Avg{q}}
,
\eea
and the phase transition is hybrid for any other $(k,n)$. 
%%$k\geq 3$, $n\geq 3$.
In 
%%Fig.~\ref{f_sim2} 
Fig.~\ref{f_2nd_hyperedge} we show the order parameter $S_{kn}$ as a function of $p_H$ for different $(k,n)$-cores for a random Poisson hypergraph with the hyperedge cardinality and node degree distributions given by Eq.~(\ref{Poisson}). The figure demonstrates excellent agreement with our theoretical predictions.

%%%%%%%%%%%%%%%%%%%%%%%%%%%%%%%%%%%%%%%%%%%%
%%%%%%%%%%%%%%%%%%%%%%%%%%%%%%%%%%%%%%%%%%%%

%%%%%%%%%%%%%%%%%%%%%%%%%%%%%%%%%%%%%%%%%%%%
%%%%%%%%%%%%%%%%%%%%%%%%%%%%%%%%%%%%%%%%%%%%

Let us now compare the equations determining the 2nd neighbor $(k,n)$-core percolation with pruning of the hyperedges  to the factor graph $(k,n)$-core equations \cite{lee2023k,mancastroppa2023hyper} characterizing the sub-factor-graph induced by the nodes of at least degree $k$ and the factor nodes of at least cardinality $n$. In this model the fraction of nodes $S_{kn}$  and the fraction of factor nodes (hyperedges) in the $(k,n)$-core $R_{kn}$ are given by 
\bea
&&
R_{kn} = p_H \sum_{m\geq n} Q(m)  \sum_{s=n}^{k} {k \choose s} W^s (1 - W)^{m-s}  
,
\nonumber
\\[3pt]
&&
S_{kn} =  p_N \sum_{q\geq k} P(q)  \sum_{s=k}^n V^s (1 - V)^{q-s}
,
\label{730b}
\eea
with $W$ and $V$ obeying
\bea
&&
V = p_H \sum_{m\geq n} \! \frac{mQ(m)}{\avg{m}} \! \sum_{s=n-1}^{m-1} \! {m-1 \choose s} W^s (1 - W)^{m-1-s}
,
\nonumber
\\[3pt]
&&
W =p_N\sum_{q\geq k} \frac{q P(q)}{\avg{q}} \sum_{s=k-1}^{q-1} {q-1 \choose s} V^s (1 - V)^{q-1-s}
.
\label{720b}
\eea

\begin{figure}[t]
%%[htbp]
\begin{center}
\includegraphics[width=\columnwidth]{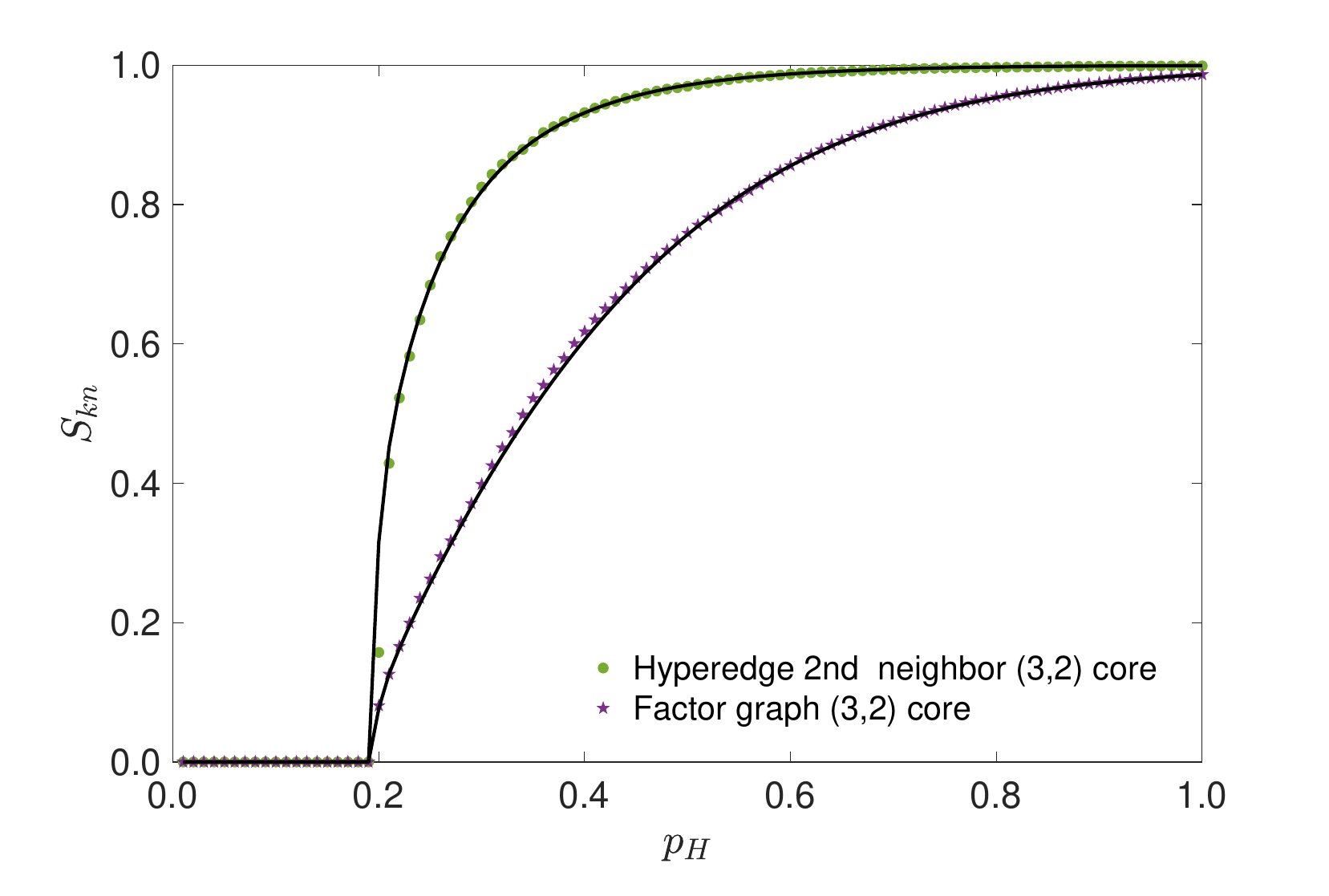} 
\end{center}
\caption{The fraction $S_{kn}$ of nodes in the $(3,2)$-core is plotted versus $p_H$ for the 2nd neighbor $(k,n)$-core with pruning of the hyperedges and for the factor graph $(k,n)$-core. The latter displays the same critical threshold of the first but a smaller fraction of nodes in the care. The hypergraph has Poisson cardinality and degree distribution with $\avg{m}=4.0$ and $\avg{q}=2\avg{m}$.}
\label{f_2nd_hyperedge_fg}
\end{figure}

We note that when $p_N=1$ the nature of the phase transition of the 2nd neighbor hypergraph $(k,n)$-core percolation and its percolation threshold  coincides with the one of the $(k,n)$-core on factor graphs. %however the equation for $S_{kn}$ differs from the corresponding equation valid in the factor graph case. 
Moreover, $R_{kn}$ coincides for the two models while $S_{kn}$ differs (see Fig.~\ref{f_2nd_hyperedge_fg}). 

%%%%%%%%%%%%%%%%%%%%%%%%%%%%%%%%%%%%%%%%%%%%
%%%%%%%%%%%%%%%%%%%%%%%%%%%%%%%%%%%%%%%%%%%%

%%%%%%%%%%%%%%%%%%%%%%%%%%%%%%%%%%%%%%%%%%%%
%%%%%%%%%%%%%%%%%%%%%%%%%%%%%%%%%%%%%%%%%%%%

%%%%%%%%%%%%%%%%%%%%%%%%%%%%%%%%%%%%%%%%%%%%
%%%%%%%%%%%%%%%%%%%%%%%%%%%%%%%%%%%%%%%%%%%%

\begin{figure}[t]
%%[htbp]
\begin{center}
\includegraphics[width=\columnwidth]{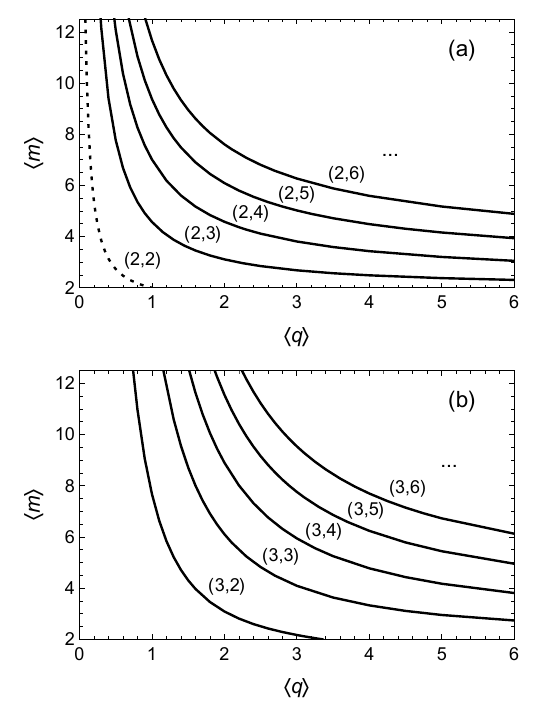} 
\end{center}
\caption{The phase diagram of the 2nd neighbor hyperedge $(k,n)$-core percolation in the $\avg{q}$--$\avg{m}$ plane is shown for  uncorrelated Poisson hypergraphs with $p_H=p_N=1$. Panel (a) displays the phase diagram for $(k,n)$-core percolation with $(k,n)$ given by  $(2,2)$, $(2,3)$, $(2,4)$, $(2,5)$, and $(2,6)$; panel (b) displays the $(k,n)$-core percolation phase diagram for $(k,n)$ given by $(3,2)$, $(3,3)$, $(3,4)$, $(3,5)$, and $(3,6)$. 
 Each core exists in the whole region to the right of the corresponding boundary. 
All boundaries are discontinuous (hybrid) transitions with one exception, namely, the $(2,2)$-core, which is always a continuous phase transition. 
For the $(2,2)$-core, the phase boundary ends at the point $\avg{q}=1$, $\avg{m}=2$. 
For the $(3,2)$-core, the phase boundary ends at the point $\avg{q}=3.3509\ldots$, $\avg{m}=2$.
}
\label{f3}
\end{figure}

%%%%%%%%%%%%%%%%%%%%%%%%%%%%%%%%%%%%%%%%%%%%
%%%%%%%%%%%%%%%%%%%%%%%%%%%%%%%%%%%%%%%%%%%%

It follows that the phase diagram of the 2nd neighbor hypergraph $(k,n)$-core percolation reduces to the phase diagram of the $(k,n)$-cores of a factor graph for $p_N=1$. 
Figure~\ref{f3} shows this phase diagram for $p_H=p_N=1$.
%%This is schematically represented on Figs.~$\ref{f3}$ and $\ref{f4}$ for $p_H=p_N=1$. 
Comparing 
%%these figures 
this figure with the corresponding one for the 1st neighbor $(k,n)$-core percolation problem, Fig.~\ref{fS}, we notice the absence of the tricritical point in Fig.~\ref{f3} with hybrid transitions only present for $k\geq 3$.

%%%%%%%%%%%%%%%%%%%%%%%%
%%%%%%%%%%%%%%%%%%%%%%%%
%%%%%%%%%%%%%%%%%%%%%%%%

The explicit equation for the continuous transition line for $(2,2)$-core  in this phase diagram is given by 
\be
\avg{m} = \frac{1+\sqrt{1+8\avg{q}^2}}{2\avg{q}}
. 
\label{00010}
\ee
Furthermore, the end point of the phase boundary (hybrid transition line) for the $(3,2)$-core is given by  
\be
\avg{q}=3.3509\ldots , \ \avg{m}=2
.
\label{00020}
\ee
Here the number $3.3509\ldots = (1 + x + x^2)/x$, where $x$ is the non-zero root of the equation:
\be
1 + x + x^2 = e^x
.
\label{00030}
\ee

\section{Conclusions}
\label{s-c}

In this work we have developed a message-passing theory 
%%, valid for locally tree-like hypergraphs 
for hypergraph $(k,n)$-core percolation  assuming that hyperedges can only be intact if all their nodes are undamaged. This simple hypothesis is relevant for a wide variety of real scenarios including supply networks, protein interactions networks and networks of chemical reactions. 
The $k$-core decomposition is a widely used tool for the discovery of highly connected substructures within complex networks, which essentially determine the character of  cooperative and spreading phenomena in networks.  {We demonstrate that  $k$-core problems for hypergraphs are significantly different from  the $k$-core problem on ordinary graphs. }  
While the hypergraph structure is represented by an equivalent bipartite graph between nodes and hyperedges---factor nodes (factor graph), here we reveal that the set of $k$-cores on hypergraphs is distinct
from this set for their 
%%corresponding 
factor graphs \cite{lee2023k,mancastroppa2023hyper}.

The reason for this difference is that the deletion of a node in a hypergraph also removes all the adjacent hyperedges, while 
the deletion of a node in a factor graph doesn't lead to
the removal of factor nodes, only the connections of the
neighboring factor nodes to the removed node disappear. 
Accounting for this difference, we describe a set of $k$-core problems (also called 1st neighbor $(k,n)$-core problems) for hypergraphs and the corresponding pruning algorithms in which nodes and hyperedges are progressively removed (damaged) if their degrees and cardinalities, respectively, fall behind given threshold values, $k$ and $n$. 
We obtain phase diagrams for such $(k,n)$-cores in random hypergraphs. In contrast to ordinary graphs, where the phase transition for the $2$-cores is continuous, while the phase transitions for $(k\geq3)$-cores are hybrid, for the $(2,2)$-core we observe two transition lines on the phase diagram---the continuous transition line and the hybrid transition one. These lines converge at the tricritical point.  

 {In order to bridge the gap between the $k$-core problems defined on hypegraphs and on factor graphs,  we introduce a novel class of hypergraph $k$-core problems, in which the pruning process involves only nodes or only hyperedges and accounts for the connectivity of their  neighbors in the factor factor graph. }We call these latter problems 2nd neighbor $(k,n)$-core percolation processes.
We show that the 2nd neighbor $(k,n)$-core percolation process where only nodes are pruned is rather distinct from the one where only hyperedges are pruned. In particular the nature of the $(k,n)$-core percolation transition and the percolation threshold of the two variants of 2nd neighbor $(k,n)$-core percolation process is different. The 2nd neighbor $(k,n)$-core percolation process with node pruning has a phase diagram that coincide with the 1st neighbor $(k,n)$-core  process.   {The 2nd neighbor $(k,n)$-core percolation process with hyperedge pruning has a phase diagram that for $p_N=1$ coincides with the factor graphs $(k,n)$-core percolation problems \cite{lee2023k,mancastroppa2023hyper}.}
Note however that the order parameters for 2nd neighbor $(k,n)$-cores with pruning of nodes/hyperedges (fractions of nodes and hyperedges within these $(k,n)$-cores) do not all reduce to the ones for the hypergraph/factor graph $(k,n)$-cores.

We suggest that this work will 
%%contribute to reveal 
highlight the important differences between hypergraphs and factor graphs and will contribute to a better understanding of specific critical phenomena in higher-order networks.   {It is our hope and trust that the first neighbor and second neighbor $(k,n)$ core hypergraph problems defined here might  find wide applications in the study or real-world higher-order networks.}
%%how higher-order networks might change the  behavior of critical phenomena unfolding on them.
%These results are obtained within a message-passing theory exact for locally tree-like hypergraphs and supported by simulations. 
%In this work we have developed a message passing theory, valid for locally tree-like hypergraphs for hypergraph $(k,n)$-core percolation  assuming that hyperedges can only be intact if all their nodes are undamaged.
%This simple hypothesis is relevant for a wide variety of real scenarios including supply networks, protein interactions networks and networks of chemical interactions. We have shown that the  $(k,n)$-core percolation problems can be formulated in terms of pruning algorithms that prune nodes and/or hyperedges on the basis of their connectivity (here also called first neighbor hypergraph $(k,n)$-core percolation or simply hypergraph $(k,n)$-core percolation). Alternatively  nodes or hyperdges can be pruned on the basis of the connectivity of their neighbors in the factor graph leading to what we  call second neighbor hypergraph $(k,n)$-core percolation problems.

%The  hypergraph $(k,n)$-core percolation is 
%distinct from the 

%%%%%%%%%%%%%%%%%%%%%%%%
%%%%%%%%%%%%%%%%%%%%%%%%
%%%%%%%%%%%%%%%%%%%%%%%%
\bibliography{k_core_references}

\end{document}